%% file: main.tex
\documentclass[sigconf,nonacm,9pt]{acmart}
\input{headers/packages}
\input{headers/jargon}

\newcommand\vldbdoi{XX.XX/XXX.XX}

\begin{document}
\title{\sys: Efficient Data Augmentation Search for AutoML}

\settopmatter{authorsperrow=4}

\author{Zezhou Huang}
\email{zh2408@columbia.edu}
\affiliation{
  \institution{Columbia University}
}
\author{Pranav Subramaniam}
\email{psubramaniam@uchicago.edu}
\affiliation{
  \institution{The University of Chicago}
}
\author{Raul Castro Fernandez}
\email{raulcf@uchicago.edu}
\affiliation{
  \institution{The University of Chicago}
}
\author{Eugene Wu}
\email{ewu@cs.columbia.edu}
\affiliation{
  \institution{DSI, Columbia University}
}

\input{sections/abstract}

\maketitle



\input{sections/introduction}

\input{sections/usecases}

\input{sections/overview}

\input{sections/factorized}
\input{sections/implementation}

\input{sections/evaluation}

\input{sections/relatedwork}
\input{sections/conclusions}

\bibliographystyle{ACM-Reference-Format}
\bibliography{main}

\end{document}

%% file: headers/packages.tex
\usepackage{xcolor}

\usepackage{sparklines}
\usepackage[export]{adjustbox}
\usepackage{xspace}
\usepackage{xcolor}
\usepackage{soul}
\usepackage{marginnote}
\usepackage{caption}
\usepackage{subcaption}
\usepackage{enumitem}
\usepackage{algorithm}
\usepackage[noend]{algpseudocode}

\usepackage{array}

\usepackage{mathtools}

\hypersetup{
  colorlinks=false,
  linkcolor=darkred,
  citecolor=darkgreen,
  urlcolor=darkblue
}

\usepackage[noabbrev,capitalise]{cleveref}

\newcounter{ex}
\newtheorem{example}[ex]{Example}
\newcounter{prob}
\newtheorem{problem}[prob]{Problem}

\usepackage[utf8]{inputenc}

%% file: headers/jargon.tex
\algnewcommand\algorithmicswitch{\textbf{switch}}
\algnewcommand\algorithmiccase{\textbf{case}}
\algnewcommand\algorithmicassert{\texttt{assert}}
\algnewcommand\Assert[1]{\State \algorithmicassert(#1)}%
\algdef{SE}[SWITCH]{Switch}{EndSwitch}[1]{\algorithmicswitch\ #1\ \algorithmicdo}{\algorithmicend\ \algorithmicswitch}%
\algdef{SE}[CASE]{Case}{EndCase}[1]{\algorithmiccase\ #1}{\algorithmicend\ \algorithmiccase}%
\algtext*{EndSwitch}%
\algtext*{EndCase}%

\newcommand{\eg}{e.g.,\ }

\newcommand{\tinyskip}{\vspace{3pt}}
\newcommand{\mypar}[1]{\tinyskip\noindent\textbf{#1.}\xspace}

\newenvironment{structure*}{\color{blue}\begin{myenumerate}}{\end{myenumerate}}

\sethlcolor{yellow}

\DeclareMathOperator*{\argmax}{argmax}

\definecolor{light-gray}{gray}{0.95}
\definecolor{mid-gray}{gray}{0.55}
\definecolor{darkred}{rgb}{0.7,0.25,0.25}
\definecolor{darkgreen}{rgb}{0.15,0.55,0.15}
\definecolor{darkblue}{rgb}{0.1,0.1,0.5}
\definecolor{blue}{rgb}{0.01,0.40,.8}

\usepackage{latexsym}

\DeclareRobustCommand{\ojoin}{\rule[0.10ex]{.3em}{.4pt}\llap{\rule[1.40ex]{.3em}{.4pt}}}
\newcommand{\leftouterjoin}{\mathrel{\ojoin\mkern-6.5mu\Join}}

\newcommand{\gray}[1]{\textcolor{mid-gray}{#1}}

\newcommand{\red}[1]{\textcolor{red}{#1}}

\newcommand{\blue}[1]{\textcolor{blue}{#1}}

\newcommand{\revise}[1]{#1}
\newcommand{\review}[1]{}

\newcommand{\sys}[0]{\textsf{Kitana}\xspace}

\newcommand{\eat}[1]{}

\newcommand{\stitle}[1]{\vspace{2pt}\noindent\textbf{#1}}

\newenvironment{myitemize}{\begin{list}{\gray{$\bullet$}}{\renewcommand{\leftmargin}{0.15in}}}{\end{list}}

%% file: sections/abstract.tex
\begin{abstract}

AutoML services provide a way for non-expert users to benefit from high-quality ML models without worrying about model design and deployment, in exchange for a charge per hour ($\$21.252$ for VertexAI). 
\revise{However, existing AutoML services are "model-centric," in that they are limited to extracting features and searching for models from initial training data---they are only as effective as the initial training data quality.}
With the increasing volume of \revise{tabular data} available, both within enterprises and to the public, there is a huge opportunity for training data augmentation. For instance, vertical augmentation adds predictive features, while horizontal augmentation adds examples. This augmented training data yields potentially much better AutoML models  at a lower cost. However, existing AutoML and augmentation systems either forgo the augmentation opportunities that provide poor models, or apply expensive augmentation searching techniques that drain users' budgets.

\sys is a "data-centric" AutoML system that {\it also} searches for new tabular datasets that can augment the tabular training data with new features and/or examples. \sys manages a corpus of datasets,  exposes an AutoML interface to users and searches for augmentation with datasets in the corpus to improve AutoML performance. To accelerate augmentation search, \sys applies aggressive pre-computation to train a {\it factorized proxy model} and evaluate each candidate augmentation  within $0.1s$. \sys also uses a cost model to limit the time spent on augmentation search, supports expressive data access controls, and performs request caching to benefit from past similar requests. 
Using a corpus of 518 open-source datasets, we show that \sys produces higher quality models than existing AutoML systems in orders of magnitude less time. Across different user requests, we find \sys increases the model R2 from ${\sim}0.16\to {\sim}0.66$ while reducing the cost by ${>}100\times$ compared to the naive factorized learning and SOTA data augmentation search.

\end{abstract}

%% file: sections/introduction.tex
\section{Introduction}
\label{sec:introduction}

Cloud-based AutoML services~\cite{Vertexai,MSautoml,oracleautoml,sagemaker} offer the promise to make machine learning (ML) \revise{over tabular data} accessible to non-ML experts. Given a \revise{tabular} training set $T$ and target variable $Y$, AutoML automatically searches for a good model. Technical barriers such as model design and tuning, training details, and even deployment barriers are hidden from the user in exchange for a charge per hour (\eg as of Nov 2022, Vertex AI~\cite{Vertexai} charges $\$21.252$). \revise{However, existing AutoML services are "model-centric": given a {\it fixed} training dataset $T$, all work is focused on transforming $T$ (e.g., by extracting features) and model search. The effectiveness of these services is constrained by the quality of  $T$. Without examples containing predictive features, extensive training will not be beneficial.}

We observe that increasingly large repositories for \revise{tabular dataset} within and across organizations (\eg data lakes, open data portals, and databases) can be used for {\it data augmentation}.  \revise{This presents an opportunity for data-centric AutoML: by efficiently searching these repositories to identify {\it new} datasets to {\it augment} the user's training data---by joining with datasets to acquire new features ({\it vertical augmentation}), unioning with datasets to acquire new examples ({\it horizontal augmentation}), or a combination of the two---we could dramatically improve AutoML quality, speed, {\it and} cost.}  At the same time, not all data in these repositories are, nor should be, accessible to everyone---even within an organization teams have varying levels of access to data.  Any augmentation system must respect these access requirements.

We believe a practical data augmentation search for AutoML should meet four criteria. (\textbf{C1}) Most importantly, augmentation search {\it must be fast}. In the cloud, time is money.  The search procedure should be independent of the augmented dataset's cardinality, avoid expensive model retraining, and efficiently search a large space of augmentation candidates.  
(\textbf{C2}) The search criteria should be correlated with model improvement.  Investing more time (and money) should return {\it better} models.
(\textbf{C3}) It should support a mixture of vertical and horizontal augmentations to make the best use of the available dataset repository. (\textbf{C4}) Finally, it should support varying levels of data release. For instance, public datasets can be shared in their raw form with the user, whereas private datasets may be shared with the augmentation system to enhance a model's predictions but {\it not} shareable with the user~\cite{Lcuyer2017PyramidES}.

Recent data augmentation systems identify a set of candidate tables for augmentation, and for each augmentation, assess its improvement to the model. Techniques like ARDA~\cite{chepurko2020arda} only support vertical augmentation (C3), materialize the augmented dataset, and retrain a model to assess the augmentation quality.  This trades-off time for model quality (C2), but is too slow and expensive for cloud environments (C1).
Other approaches such as Li et al.~\cite{li2021data} use a proxy ``Novelty'' metric to estimate only horizontal augmentation candidates (C3). However, our experiments show that ``Novelty'' is both slow (C1) and can be uncorrelated with actual model improvements (C2). 
To summarize, existing approaches are slow (C1), may not actually improve the model quality (C2), do not support a mixture of horizontal and vertical augmentation (C3), and do not support different levels of data release (C4).

\begin{figure}
    \centering
    \includegraphics[width=1.\columnwidth]{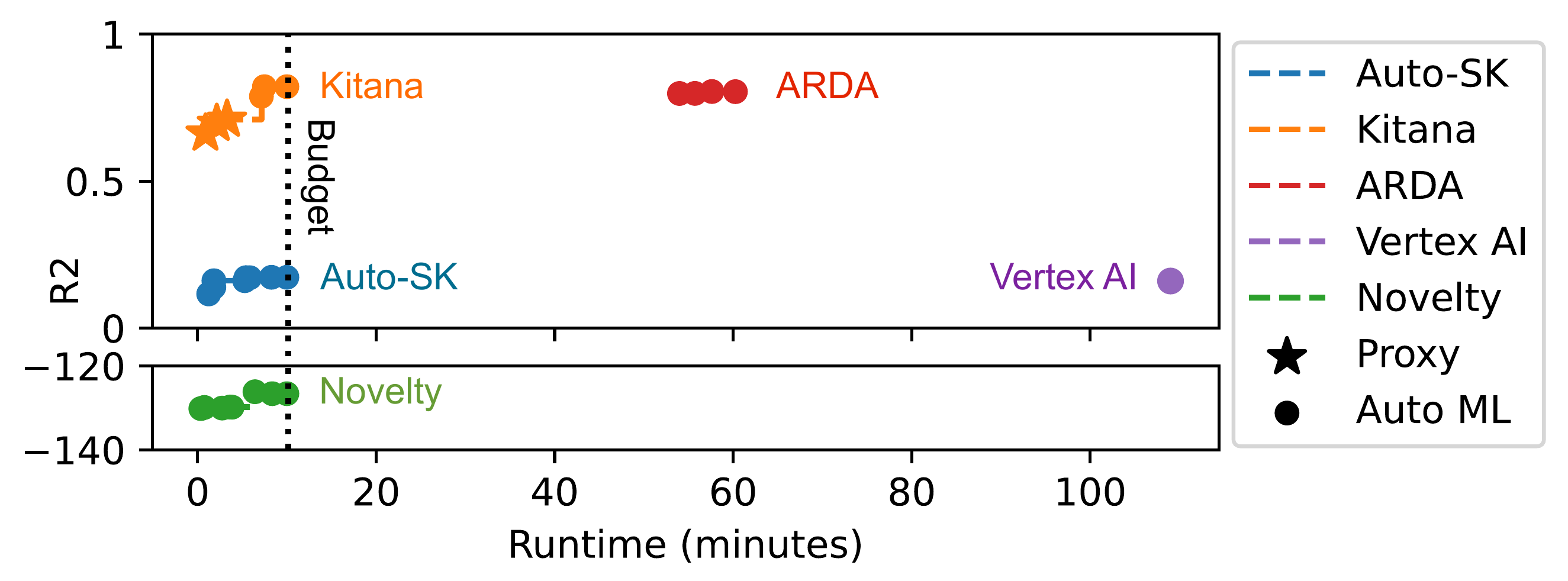}
    \vspace*{-8mm}
    \caption{AutoML performance (testing $R2$) for $Gender$ dataset. The time budget is 10 minutes (dotted vertical line). \sys allocates $\sim3$ minutes for data augmentation using a corpus of 517 datasets with a proxy model and reaches $R2=\sim0.7$; \sys then sends the augmented dataset to AutoML and further improves $R2$ to $0.82$ with the remaining budget. Other baselines are either slow or have low $R2$.
    }
    \label{fig:moneyfig}
\end{figure}

This paper describes \sys, a data-centric AutoML system that satisfies the above four criteria.  \sys combines and extends ideas from factorized learning~\cite{schleich2016learning}
and indexing~\cite{fernandez2018aurum} to enable the following novel capabilities.   We introduce a cheap-to-train \emph{proxy model} that  reflects the final model quality and is designed to avoid expensive model retraining for each augmentation candidate.  
Specifically, we pre-compute cheap sketches for each dataset using ideas from factorized learning and semi-ring aggregation, and use them to efficiently update the proxy model (a linear model) in $O(n^2j)$ time\footnote{Where $n$ is the number of features and $j$ is the join key domain size; note the time complexity is independent of the cardinality of the augmented dataset} for any augmentation.   

Given a time budget $t$, \sys splits it between augmentation search and AutoML training. \sys estimates the model search time~\cite{popescu2013predict,wang2015performance}: when the estimated time exceeds the remaining budget, \sys stops augmentation search early and trains the final model using an existing AutoML service or library. This helps \sys stay within budget {\it and} return a model that is more accurate than all existing data augmentation and AutoML systems.

\Cref{fig:moneyfig} summarizes our major results compared to prior data augmentation techniques (ARDA~\cite{chepurko2020arda}, Novelty~\cite{li2021data}); AutoML libraries (Auto-sklearn); and cloud AutoML services (Google's Vertex AI). Since augmentation techniques don't accept a time budget and return the augmented dataset, we run them to completion and then run Auto-sklearn to derive a model.
\revise{Both Auto-sklearn and Vertex AI only take the original training dataset as input, thereby neglecting the abundant augmentation opportunities that may be present in the data corpus.}
We set Auto-sklearn's budget to 10 minutes and Vertex AI's budget to 1 hour (minimum allowed), and report $R2$ throughout the model search; Vertex AI doesn't enforce the budget.  

The training set doesn't have predictive features, so Auto-sklearn and Vertex AI do not find good models.
ARDA takes ${\approx}50$ minutes for data augmentation, and the final model is still slightly worse than \sys ($R2=0.81$ vs $0.82$). Novelty is uncorrelated with model accuracy and degrades the final model. Finally, \sys's  proxy models (stars) are already highly competitive, and it leaves enough time to train the final model using AutoML (circles).  

In addition to higher accuracy, faster results, and cheaper cost, \sys supports sensible dataset access controls via \emph{data release labels}. Labels indicate what can be released to the user, and what data providers need to share with \sys. A data provider can label each dataset so the raw data can be released to the user ({\it RAW}), only the trained model can be released ({\it MD}), or the user can perform inference using the model but the model cannot be released ({\it API}). A user request can specify whether they want the full augmentation plan and associated datasets, a model trained on augmented data, or a prediction API, and \sys will automatically adjust its search procedure to satisfy data release labels. Although the implementation to support these controls is relatively straightforward, the labels mirror access control expectations in the wild~\cite{Lcuyer2017PyramidES,dourish2004security}.  They also illustrate the complex interaction between dataset access, system design, and user-facing API semantics. We leave a more complete treatment of access control to future work.

We evaluate \sys using a corpus of 518 datasets collected from open data repositories and compare its performance against existing AutoML services like Auto-Sklearn and Google's Vertex AI. Across all user requests based on 5 real-world datasets, we find that \sys is faster, cheaper, and of higher quality than AutoML services.   Specifically, \sys takes less than 1 minute to exceed the accuracy that AutoML reaches after ${\ge}10min$; for the same time budget, \sys increases the model $R2$ scores by ${\sim}0.05{-}0.30$.

\smallskip\noindent To summarize, we contribute the following:
\begin{myitemize}
  \item The design and implementation of \sys, an AutoML system that uses practical data augmentation. \sys efficiently searches a corpus of datasets for augmentation opportunities that increase the final model as compared to using AutoML services alone.  
  \item \sys leverages careful pre-computation and factorized learning to evaluate each augmentation within ${\sim}100ms$.
  \item Extensive evaluation against two AutoML systems (Google's Vertex AI~\cite{Vertexai} and Auto-sklearn) using a corpus of real-world datasets. We show that \sys is cheaper and better: \sys reduces request latency by over $10\times$ for the same model accuracy and increases accuracy by over $30\%$ for the same time budget.
\end{myitemize}

%% file: sections/usecases.tex
\section{Problem Description}
\label{sec:probd}

This section introduces our terminology, related AutoML services, and the augmentation search problem definition.

\subsection{Preliminaries}

\sys focuses on supervised ML tasks over tables.

\mypar{Tables and Dataset Corpus} Let $D$ denote a general relational table and the list of attributes $S_D = [a_1,\ldots,a_m]$ be its schema.  
A dataset corpus $\mathcal{D} = \{D_1,...,D_n\}$ is a set of tables.
We focus on two forms of augmentation: {\it horizontal augmentation} unions two tables $D_1 \cup D_2$ that are schema compatible, and  {\it vertical augmentation} joins two tables $D_1 \Join D_2$ together to add additional attributes.

\mypar{Supervised ML} In supervised ML tasks like classification and regression, a model with parameters $\theta$ is trained on a tabular dataset $T$ with features $X\subset S_T$ and target $Y\in S_T$. The goal is to predict $Y$ for unseen $X, Y$ pairs. Users may specify the model type $M$ or use a meta-search process. Model performance is evaluated using cross-validation~\cite{refaeilzadeh2009cross} during training and hyperparameter optimization. The final returned model is evaluated once on a test dataset $V$.

\mypar{AutoML}
Today's AutoML services input $T$ and $Y$,
automatically searching for suitable data transformations and model parameters $\theta$. Some also deploy models and provide inference APIs
This allows non-experts in ML to benefit from model training and deployment without worrying about the details. Developers can use  open-source packages like Auto-sklearn~\cite{feurer2020auto} and FLAML~\cite{wang2021flaml}, and commercial solutions like  H2O~\cite{h2o}.  End users can use cloud-based services like Google Vertex AI~\cite{Vertexai}, Microsoft AutoML~\cite{MSautoml}, Amazon SageMaker~\cite{sagemaker}, and Oracle AutoML~\cite{oracleautoml}.

Model training and search are time-critical as the services use hourly pricing: for instance, as of November 2022, Vertex AI charges  $\$21.252$/hr on tabular data, Amazon Sagemaker charges  $\$0.115-5.53$/hr, and Microsoft AutoML charges $\$0.073-27.197$/hr.  Running AutoML longer increases the search space, enhancing the chances of finding accurate models. 
Since the search can take a long time,  users typically set a time budget $t$.  

More data is available than ever before---through public government portals, enterprise data lakes, data marketplaces, logging tools, and IoT devices.   This is a huge opportunity to go beyond merely model search ~\cite{sambasivan2021everyone} for AutoML, and to also vertically and horizontally augment training datasets~\cite{li2021data,chepurko2020arda}.

\subsection{Data Augmentation for Tabular AutoML}
\label{sec:usecase}

We present  use cases to motivate \sys.

\revise{\stitle{Public Health.} Consider a public health analyst working with a Manhattan health dataset, which contains features about the patient's residence, and aims to predict asthma symptoms~\cite{school-buses-improve}. The AutoML-trained model has low accuracy due to the absence of predictive features and examples.
She submits her training dataset to \sys, which quickly finds neighborhood-level datasets about air quality from NYC Open Data~\cite{nycopen}, restaurant hygiene from Yelp, and additional asthma symptom records about Brooklyn and the Bronx that a colleague had registered.  She approves the use of all three datasets, and her model improves considerably.  }

\revise{\stitle{Churn Prediction.} Sales analysts are trying to use AutoML to model customer churn. The initial datasets only contain customer demographic (age and income) and product features (purchase date and price) collected by their own team, which are not predictive. They submit the task to \sys, which searches the company's data warehouse used by teams throughout the company. 
\sys finds two datasets that greatly improve model performance: a customer engagement dataset containing website visits, which shows how actively customers interact with the company, and a dataset of local unemployment rates, which reflects customers' purchasing power. 
The analysts contact the owners of these datasets to gain direct access, and use them to better identify customers likely to churn. }

\begin{figure}
    \centering
    \includegraphics[width=0.8\columnwidth]{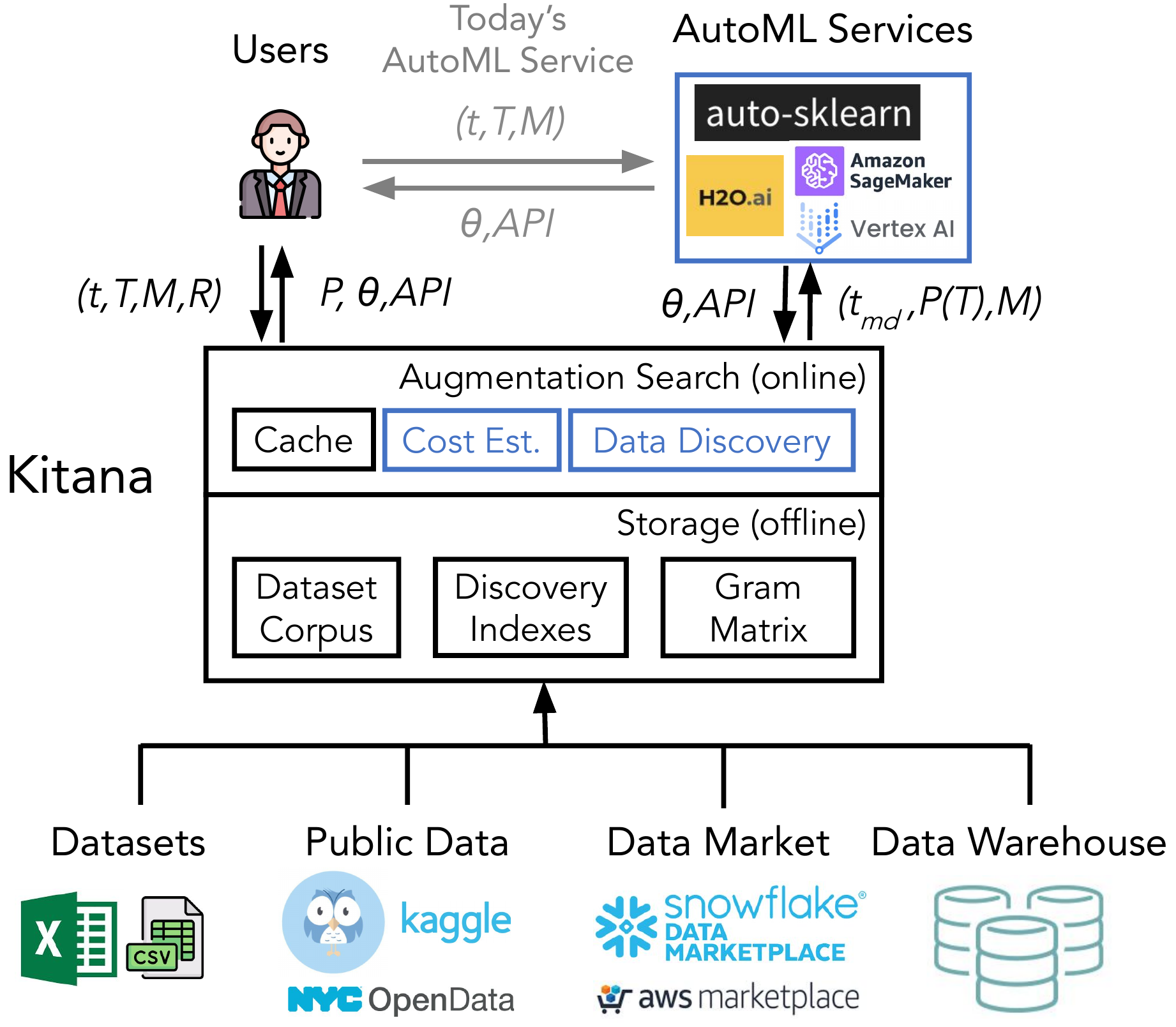}
     \vspace*{-5mm}
    \caption{Overview of \sys Architecture.}
    \label{fig:service_architecture}
\end{figure}

\subsection{Problem Definition}\label{ss:problem}

Let $\mathcal{D}$ be a corpus of datasets.
Each dataset $D{\in}\mathcal{D}$ is assigned an access label $l(D){\in}\{RAW,MD,API\}$ from data providers for access control.
The labels are  ordered $RAW{<}MD{<}API$: 
$RAW$ allows access to $D$. $MD$ allows access to models $\theta$ trained over $D$, but not $D$ itself. $API$ disallows access to $\theta$, but allows a prediction API that uses $\theta$.

The user submits a tuple $(t,T,M,R)$, which consists of a time budget $t$, training dataset $T$,  an optional model type $M$ ($l$ for linear regression, $\cup$ for all models), and return labels $R{\subseteq}\{RAW,MD,API\}$ that describe what \sys should return
$RAW$ for raw augmentation data with the augmentation plan introduced next, $MD$ for model $\theta$, and $API$ for prediction API. User also has a testing dataset $V$ for evaluation, but it's only accessible by the user.

We search for a good {\it Augmentation Plan} $\mathcal{P} = [A_1,\cdots, A_n]$ that specifies a sequence of horizontal or vertical augmentation $A$: Horizontal augmentation contains augmentation data $D_A$ and vertical augmentation additionally contains join key $j_A$. $A(I)$ applies augmentation to dataset $I$,  where $A(I) = I \cup D_A$ for horizontal augmentation and $A(I) = I \Join_{j_A} D_A$ for vertical augmentation\footnote{For notational convenience, the text assumes the $j_A$ is the same in $I$ and $D_A$.  More generally, \sys uses Aurum for candidate discovery, which supports equijoins; the factorized learning techniques are general to any join, including theta and anti-joins.}.  Applying $\mathcal{P}$ to training set $T$ is defined as $\mathcal{P}(T) = A_n(A_{n-1}(\cdots A_1(T)))$.

The access label $l(D)$ and the requested return labels $R$ limit the search space to $\sigma_{l(D)\leq min(R)}(\mathcal{D})$.  
When $min(R){\geq}MD$, only horizontal augmentation is allowed because the user can't vertically augment new features for prediction using $\theta$. $R$ therefore provides a trade-off between performance and explainability: $R{=}\{API\}$ allow all $\mathcal{D}$, but to inspect raw data and model ($min(R){=}RAW$), the search is over a restricted subset.

Next, we present the formal problem definition:

\begin{problem}
\label{mainproblem}
  Given request $(t, T, M, R)$ and testing dataset $V$ (not available during search), find an augmentation plan $\mathcal{P^*}$ that maximizes the testing accuracy:
\begin{flalign*}
\mathcal{P}^* = &\quad\argmax_{\mathcal{P}}   acc(\theta, \mathcal{P}(V))&&\\
s.t. &\quad\theta = AutoML.train(\mathcal{P}(T)), &&  \\
&\quad t_{data} + t_{md} \le t, && \text{(time budget constraint)}\\
&\quad\theta \in M,   && \text{(model type constraint)}\\
&\quad\forall D \in \mathcal{P}^*, l(D) \leq min(R). && \text{(access control constraint)}
\end{flalign*}
  where $t_{data}$ and $t_{md}$ are the times for augmentation search and AutoML model search, respectively.
\end{problem}

\stitle{Hardness.} Feature selection is NP-hard~\cite{chen1997minimum} that can be reduced to finding the optimal augmentation plan, where each feature is a normalized vertical augmentation. In practice, applying AutoML to evaluate every augmentation plan is expensive and not scalable.

%% file: sections/overview.tex
\section{\sys Overview}
\label{sec:overview}
We now present the architecture, how \sys wraps the problem definition to provide AutoML service, and user request lifecycles.

\subsection{\sys Architecture}
\label{subsec:arch}

\Cref{fig:service_architecture} presents the system components and control flow, where the black components are part of \sys and the blue ones can be outsourced. \Cref{sec:imp} describes the components in greater detail.

Existing AutoML services follow the gray path, where the user submits a request and receives model parameters and/or a prediction API.  \sys, however, acts as an intermediary between users and AutoML.  During online operation, \sys finds the optimal augmentation plan $\mathcal{P}^*$ (\Cref{mainproblem}), passes the augmented dataset to AutoML.  Depending on the request's return labels, \sys returns the trained model $\theta$, prediction API, and/or augmentation plan $\mathcal{P}^*$.

To search for $\mathcal{P}^*$ efficiently, \sys uses the bottom-up feature selection~\cite{guyon2003introduction} to
greedily adds the next best augmentation.  It employs {\it proxy models} for evaluation to avoid retraining.
Then, if the desired model is linear $M=l$, \sys returns the {\it proxy model}. 
Otherwise, it materializes and sends the augmented dataset to AutoML.
During offline setup, data providers register datasets, and \sys builds sketches and indexes to speed up augmentation search.

\stitle{Offline Phase.}
Data providers call \sys's $upload(D, l)$ function to upload dataset $D$ with access level $l$ (\Cref{ss:problem}). \sys then computes sketches for data discovery and augmentation search---it pre-aggregates attribute profiles for data discovery systems (\sys uses Aurum~\cite{fernandez2018aurum}, see \Cref{sec:preprocess}) and factorized sketches to accelerate  augmentation candidate assessment (\Cref{sec:factorized}). \sys also applies simple cleaning, standardization, and feature transformations. 
Optionally, the data provider could upload the profiles and sketches, so \sys does not have access to the raw datasets.  

\stitle{Online Phase.} 
The user submits request $(t,T,M,R)$, and the {\it Augmentation Search} component first checks the cached augmentation plan $\mathcal{P}'$ from prior requests with the same training schema $S_T$. 
It then uses data discovery (Aurum~\cite{fernandez2018aurum} in our implementation) to find augmentable (union- or join-compatible) datasets; Aurum takes attribute profiles of training data $T$ as input (e.g., MinHash, covariance matrix), and returns a set of candidate augmentations. After that, \sys evaluates each augmentation candidate via the {\it proxy model} and greedily builds the augmentation plan. \sys further combines {\it proxy model} and factorized learning with pre-processing to reduce the evaluation cost to ${\sim}100ms$ per dataset (\cref{sec:factorized}).

\sys needs to judiciously use the time budget, especially if there are no good augmentations.   To do so, \sys uses the sketches to compute the output size of an augmentation candidate.  It uses the size to estimate the time to run AutoML $t_{md}$ on the augmented dataset using existing estimation techniques~\cite{popescu2013predict,wang2015performance,scitime}. 
If the augmentation candidate does not improve the proxy model's cross-validation accuracy, or when the estimated AutoML cost exceeds the remaining budget (\Cref{costestimation}), \sys materializes the augmented training dataset $T' = \mathcal{P}^*(T)$ and calls the AutoML service.

Finally, based on user-requested return types $R$, \sys returns the augmentation plan $\mathcal{P}^*$ along with raw augmentation datasets $D$, model $\theta$, and/or the prediction API. The API takes non-augmented records $T'$ as input, applies $\mathcal{P}^*(T')$, makes a prediction using $\theta$, and returns the prediction to the user.  

\subsection{Request Processing Flow}
\label{subsec:algo}
\input{sections/marketalg.tex}

\Cref{a:handlereq} outlines the core logic in \sys when handling a request. The user sends a request $(t,T,M,R)$ and \sys outputs $API, \theta,\mathcal{P}$ based on user-requested returned types $R$ (\Cref{ss:problem}).

To start, \sys initializes an empty augmentation plan $\mathcal{P}$.
If $T$ is schema compatible with a prior request, \sys evaluates the cached plans $\mathcal{P}'$ and chooses a plan that most improves the proxy model performance by $\ge\delta$ (L2,3). 
\sys then greedily builds the augmentation plan until the estimated time to materialize and train the model exceeds the remaining time budget (L4-16).

In each iteration (L4), \sys uses data discovery
to find a set $\mathcal{A}$ of horizontal and vertical augmentation candidates that are also compatible with the request's return types and the datasets' data release labels (L6).
Each augmentation candidate $A$ specifies the augmentation $type$ (``horiz'' or ``vert''), the annotated relation $D$ (\Cref{sec:factorizedbackground}), and join key $j$ for vertical augmentations (L8). To simplify the search procedure, \sys applies horizontal before vertical augmentation (L9).  
\revise{The intuition is that vertical augmentations done before horizontal ones can also occur after, but the reverse isn't always true because vertical augmentation alters the schema, thus affecting the unionability.
Occasionally, applying vertical augmentation first might yield better results. For instance, when the vertically augmented feature matches one feature in horizontal augmentation, and that in horizontal augmentation has a higher quality.  Nonetheless, such instances are uncommon, and we did not observe them in experiments. \Cref{a:handlereq} can be adapted to explore both horizontal and vertical augmentation by removing L9.}

Before evaluating candidate $A$, \sys adds it to the current plan $\mathcal{P^*}$(L10) and  cheaply estimates the shape (\# of attributes and rows) of augmented training data (L11) by an optimized count query (\Cref{sec:factorizedbackground}). If AutoML is needed ($M \neq l$), \sys uses a cost model to estimate how long AutoML would take on it---if the estimate exceeds the remaining time budget $t_{remain}$, \sys skips the candidate (L12). In addition, a separate timer thread terminates the augmentation search if the estimated time to run AutoML on the best augmentation plan so far may exceed the remaining budget.  

 To evaluate candidate $A$, \sys trains the proxy model and evaluates the model using 10-fold cross-validation (L13). Internally, these two steps are factorized: $T'$ in L11 is not materialized and only the aggregates needed for training and evaluation in L13 are computed.  Our pre-processing optimization further accelerates this evaluation (\Cref{sec:fac_aug}), and \sys keeps the candidate if it improves the best accuracy so far (L14). After evaluating the candidate plan, \sys keeps it if the accuracy improvement is $\ge\delta$ (L14) and there is time to train the augmented dataset $\mathcal{P^*}(T)$ using AutoML (L15).    Otherwise, it breaks the loop.

If the user requests models other than linear regression, \sys materializes the final augmented training dataset $\mathcal{P^*}(T)$ and sends it along with the remaining budget $t_{remain}$ and model types $M$ to the AutoML (L17). Otherwise, \sys skips this step and uses the proxy model.  \sys uses the final model parameters (from AutoML or proxy model) to build an API that first applies $\mathcal{P^*}$ then the model for prediction (\Cref{sec:apiconstruct}); if  AutoML returns APIs, \sys can wrap them as well, however the user will incur the AutoML service's costs for inference.
\sys then caches the new plan (L18). 
Finally, depending on $R$, \sys returns the augmentation plan (along with the raw datasets), trained model $\theta$, and/or new prediction API (L19).

%% file: sections/marketalg.tex
\algblock{Input}{EndInput}
\algnotext{EndInput}

\begin{algorithm}[b]
  \caption{HandleRequest($t,T,M, R$)}
  \label{alg-join}
  \begin{algorithmic}[1]
    \State $\mathcal{P^*} =$ initEmptyPlan() \Comment{\gray{Empty augmentation plan}}
    \State $\mathcal{P'} =$ cacheLookup(T)

    \State $\mathcal{P^*} = \mathcal{P'}$ \textbf{If} $\mathcal{P'}$ increases accuracy $\ge\delta$
    
    \While{true} \Comment{\gray{Greedily add augmentations}}
    \State $(acc^*, A^*) = (-\infty, \emptyset)$  \Comment{\gray{Best accuracy, augmentation found}}
    \State $\mathcal{A} $= dataDiscovery($\mathcal{P^*}(T).profile, R$)
    \For{$A \in \mathcal{A}$ } 
    \State $type, D, j = A$ \Comment{\gray{Aug. type, data, (opt.) join key}}
      \State \textbf{continue} \textbf{If} type $=$ ``horiz'' $\land$ $\mathcal{P^*}$ has vert. aug
      \State $\mathcal{P}' = \mathcal{P^*}.add(A)$ \Comment{\gray{Add augmentation to the plan.}}
      \State $T' = \mathcal{P}'(T)$ \Comment{\gray{Augment data (not materialized).}}
      \State \textbf{continue} \textbf{If} $M \neq l \wedge t_{remain} < cost(T', M)$
      
      \State $acc, \theta = model.trainAndEvaluate(T')$
      \State $(acc^*, \theta^*, A^*, \mathcal{P^*}') = (acc, \theta, A, \mathcal{P}')$ \textbf{If} $acc>acc^*$
    \EndFor

    \State\textbf{break} \textbf{If} $\Delta acc^*<\delta \vee (M\neq l\wedge t_{remain}<cost(\mathcal{P^*}'(T),M))$
    \State $\mathcal{P^*} = \mathcal{P^*}'$
    \EndWhile
      \State $\theta^* =$ AutoML($t_{remain}, \mathcal{P^*}(T).materialize(), M$) \textbf{If} $M \neq l$
    \State SaveToCache(T, $\mathcal{P^*}$)
    \State \textbf{return} $\mathcal{P^*}$ \textbf{If}$RAW {\in} R$; $\theta^*$ \textbf{If}$MD {\in} R$;  $buildAPI(\theta^*, \mathcal{P^*})$ \textbf{If}$API {\in} R$
  \end{algorithmic}
  \label{a:handlereq}
\end{algorithm}

%% file: sections/factorized.tex
\section{Factorized Data Augmentation}
\label{sec:factorized}

Factorized learning~\cite{khamis2018ac,schleich2019layered,cjt}  trains ML models over the output of a join query $\mathbf{X} = \mathbf{R}_1 \Join ... \Join \mathbf{R}_k$ without materializing the join.
It treats  training as an aggregation function and distributes the aggregation through the joins.  As such, the cost is linear in the relation sizes rather than the join results.
Factorized learning has been extended to many models, including linear regression~\cite{schleich2016learning}, ridge regression with regularization, factorization machines~\cite{schleich2019layered}, classification models using ridge regression~\cite{peng2020discriminative}, generalized linear models~\cite{huggins2017pass}, K-means~\cite{curtin2020rk}, and SVMs~\cite{khamis2020functional}.

Our primary contribution is to show that factorized learning is a natural fit for data augmentation.
Evaluating an augmentation candidate---materializing the augmented dataset, retraining the model, and cross-validation---is by far the biggest bottleneck~\cite{chepurko2020arda,li2021data} in \sys, and factorization avoids materialization altogether.
Thus, \sys trains and cross-validates a factorized model as the proxy to evaluate an augmentation candidate, and uses linear regression by default.
In fact, proxy models are often more accurate than AutoML models on the user's original training data (without augmentation), take orders of magnitude less time, and are often preferred in practice for interpretability and reliability reasons~\cite{xin2021production}.

However, naively using factorized learning still needs to repeatedly compute these ``training'' aggregates for each augmentation candidate. Instead, \sys identifies aggregates sharable across augmentations and pre-computes them---they are computed independently for each dataset in the repository, and in theory could be offloaded to the data provider during dataset registration.
This lets \sys outperform naive factorized learning by over two orders of magnitude on augmentation benchmarks. The rest of this section first introduces the basic concepts in factorized learning, then describes the working sharing opportunities \sys exploits for efficient data augmentation search.

\subsection{Basic Concepts}
\label{sec:factorizedbackground}
We now present semi-ring annotations and aggregation push-down, which encapsulate the core factorized learning optimizations. We then illustrate these concepts using linear regression as an example.

\subsubsection{Semi-ring Annotations}
A semi-ring is a set $\mathbf{S}$ that contains 0 and 1 elements, and supports the binary operators $+$ and $\times$; it is typically defined as the tuple $(\mathbf{S}, 0, 1, +, \times)$.  Different semi-rings are designed to support different model types.   Each tuple is annotated with an element from $\mathbf{S}$, and $R[t]\in\mathbf{S}$ denotes tuple $t$'s annotation in relation $R$.
Relational operators are extended to compute their output relation's annotations: $\Join$ multiplies the  annotations of matching input tuples, while $\gamma$ sums the  annotations in the same group-by bin. Formally, given relations $R,T$ with schemas $S_R,S_T$:
\begin{align}
  (R\Join T)[t] =&  R[\pi_{S_R} (t)] \times T[\pi_{S_T} (t)] \\
(\gamma_\mathbf{A} R)[t] = & \sum \{R[t_1] |  t_ 1 \in R , t = \pi_{\mathbf{A}} (t_1)\}
\end{align}
Here, $t$ is the output tuple, and $R[\pi_{S_R}(t)]$ is the input annotation of $t$ projected onto $R$.   

\begin{figure}
  \centering
    \hfill
    \begin{subfigure}[b]{0.1\textwidth}
         \centering
         \includegraphics[width=\textwidth]{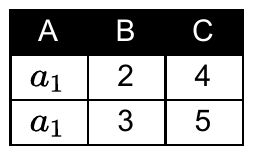}
         \caption{$\mathbf{R}_1$}
         \label{fig:r1}
     \end{subfigure}
     \hfill
     \begin{subfigure}[b]{0.1\textwidth}
         \centering
         \includegraphics[width=\textwidth]{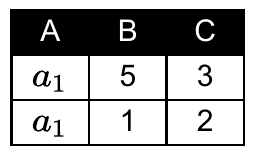}
         \caption{$\mathbf{R}_2$}
         \label{fig:r2}
     \end{subfigure}
    \hfill
    \begin{subfigure}[b]{0.075\textwidth}
         \centering
         \includegraphics[width=\textwidth]{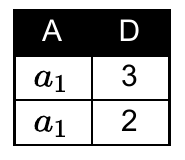}
         \caption{$\mathbf{R}_3$}
         \label{fig:r3}
     \end{subfigure}
    \hfill
     \begin{subfigure}[b]{0.1\textwidth}
         \centering
         \includegraphics[width=\textwidth]{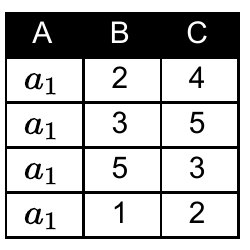}
         \caption{$\mathbf{R}_1 \cup \mathbf{R}_2$}
         \label{fig:union}
     \end{subfigure}
     \hfill
         \begin{subfigure}[b]{0.21\textwidth}
         \centering
         \includegraphics[width=\textwidth]{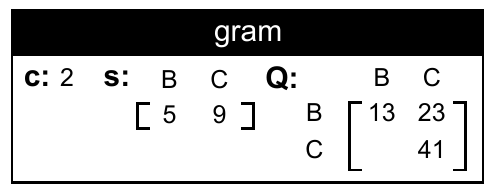}
         \caption{$\gamma(\mathbf{R}_1)$}
         \label{fig:r1_gram}
     \end{subfigure}
     \hfill
         \begin{subfigure}[b]{0.21\textwidth}
         \centering
         \includegraphics[width=\textwidth]{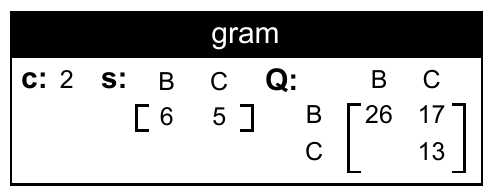}
         \caption{$\gamma(\mathbf{R}_2)$}
         \label{fig:r2_gram}
     \end{subfigure}
     \hfill
     \begin{subfigure}[b]{0.25\textwidth}
         \centering
         \includegraphics[width=\textwidth]{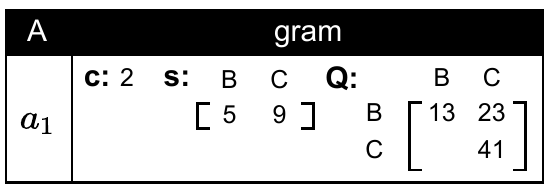}
         \caption{$\gamma_A(\mathbf{R}_1)$}
         \label{fig:r1_A}
     \end{subfigure}
     \hfill
    \begin{subfigure}[b]{0.25\textwidth}
         \centering
         \includegraphics[width=\textwidth]{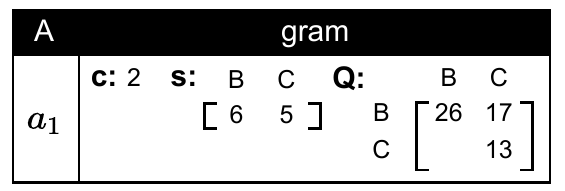}
         \caption{$\gamma_A(\mathbf{R}_2)$}
         \label{fig:r2_A}
     \end{subfigure}
     \hfill
     \begin{subfigure}[b]{0.19\textwidth}
         \centering
         \includegraphics[width=\textwidth]{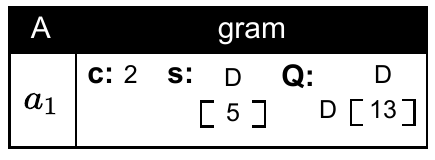}
         \caption{$\gamma_A(\mathbf{R}_3)$}
         \label{fig:r3_gram}
     \end{subfigure}
     \hfill
     \begin{subfigure}[b]{0.215\textwidth}
         \centering
         \includegraphics[width=\textwidth]{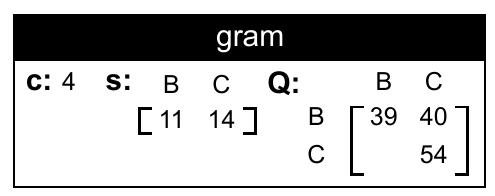}
         \caption{$\gamma(\mathbf{R}_1 \cup \mathbf{R}_2)$}
         \label{fig:union_gram}
         \vspace*{-6mm}
     \end{subfigure}
     \hfill
     \begin{subfigure}[b]{0.255\textwidth}
         \centering
         \includegraphics[width=\textwidth]{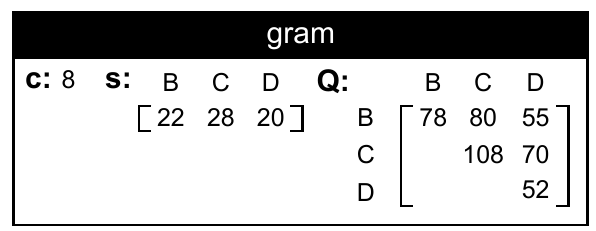}
         \caption{$\gamma((\mathbf{R}_1 \cup \mathbf{R}_2)\leftouterjoin_A \mathbf{R}_3)$}
         \label{fig:join_gram}
         \vspace*{-6mm}
     \end{subfigure}
     \hfill
  \caption{Factorized Learning Example}
  \label{fig:factorizedex}
\end{figure}

\begin{example}
Consider aggregation query $\gamma_{A, SUM(B)}(R_1\Join R_3)$ (\Cref{fig:r1} and \Cref{fig:r3}).
To support $SUM(B)$ aggregation, we use natural number semi-ring, and initially annotate $R_1[t] = \pi_B(t), R_3[t] = 1$. $R_1\Join R_3$ will produce a cartesian product (as $A$ has the same value) of four tuples with annotations $2\times1, 3\times1,2\times1, 3\times1$. $\gamma_{A, SUM(B)}$ then sums the annotations: for the only $a_1$ group, the sum is $2+3+2+3=10$.
\end{example}

The core factorized learning optimization is to push aggregation through joins by exploiting the fact that join distributes over addition as in elementary algebra.  For the previous example, we can rewrite $\gamma_{A, SUM(B)}(R_1\Join R_3)=(\gamma_{A,SUM(B)}(R_1))\Join_A(\gamma_{A,SUM(B)}(R_3))$; the sum for group $a_1$ then is computed as $(2+3)\times(1+1)=10$.
More generally, assuming $j$ is the set of join keys between $R$ and $T$:
\begin{align*}
  \gamma_\mathbf{A}\left(R \Join_j T\right) 
=  \gamma_\mathbf{A}\left(\gamma_{\mathbf{A},j}(R)\Join_j\gamma_{\mathbf{A},j}(T) \right) 
\end{align*}

\noindent This reduces intermediate sizes and thus join-aggregation costs.

\subsubsection{Factorized Learning}

We now illustrate how linear regression can be expressed as a single semi-ring join-aggregation query.

Linear regression uses training data $\mathbf{X}$ and target variable $Y$ to learn parameters $\theta$ that minimize the model's squared loss: $\hat{\theta} = arg\,min_{\theta}||Y - \mathbf{X} \theta||^2$. The closed form solution is $\theta=(\mathbf{X}^T\mathbf{X})^{-1}\mathbf{X}^TY$.  By treating $Y$ as a special feature, we can see that the bottleneck computation is the gram matrix $\mathbf{X}^T\mathbf{X}$, where each cell is the sum of products between pairs of features. $\mathbf{X}^T\mathbf{X}$ has shape $m\times m$, where $m$ is the number of features and is assumed to be much smaller than the number of rows. As a result, computing $\theta$ could be considered as a cheap postprocessing over the small $\mathbf{X}^T\mathbf{X}$.

The gram matrix semi-ring~\cite{schleich2016learning} helps efficiently compute $\mathbf{X}^T\mathbf{X}$. For a training set with $m$ features, the gram matrix annotation is a triple $(c,\mathbf{s},\mathbf{Q})\in(\mathbb{Z},\mathbb{R}^m, \mathbb{R}^{m\times m})$ that respectively contain the tuple count, sum of features, and sum of products between each pair of features. The zero and one elements are $\mathbf{0} = (0,\mathbf{0}^m, \mathbf{0}^{m\times m})$ and $\mathbf{1} = (1,\mathbf{0}^m, \mathbf{0}^{m\times m})$. The  $+$ and $\times$ operators between two annotations $a = (c_a,\mathbf{s}_a, \mathbf{Q}_a)$ and $b = (c_b,\mathbf{s}_b, \mathbf{Q}_b)$ are defined as:
\begin{align}
a + b =&  (c_a + c_b,\mathbf{s}_a + \mathbf{s}_b, \mathbf{Q}_a + \mathbf{Q}_b) \\
a \times b =& (c_a c_b,c_b\mathbf{s}_a + c_a\mathbf{s}_b, c_b\mathbf{Q}_a + c_a\mathbf{Q}_b + \mathbf{s}_a \mathbf{s}_b^T +\mathbf{s}_b \mathbf{s}_a^T)  
\end{align}
All rules for combining annotations remain the same. 

Finally, computing $\mathbf{X}^T\mathbf{X}$ is reduced to executing a single join-aggregation query $\gamma(\mathbf{R}_1 \Join ... \Join \mathbf{R}_k)$, where aggregation can be pushed down before join as discussed before.

\subsubsection{Factorized Model Evaluation}
Given the model parameters $\theta$, the semi-ring annotations of the validation table also help accelerate cross-validation.   Suppose the evaluation metric is the squared loss: $\sum (y - \theta x)^2 = \sum (y^2 - 2\theta x y + \theta^2 x^2) = \sum y^2 - 2\theta \sum x y + \theta^2 \sum x^2$. This expression decomposes into the sums of pairwise products, which are readily available in the gram matrix semi-ring annotations.

\subsection{Factorized Data Augmentation}
\label{sec:fac_aug}
Although factorized learning quickly trains a single model in one query, augmentation search still needs to execute $O(KU(N+M))$ training queries, where there are $U$ user requests, each final augmentation plan contains $K$ augmentations, and the data discovery service respectively returns $M$ and $N$ horizontal and vertical augmentation candidates. For instance, for an AutoML service with $U=1000$ user requests over $M+N=500$ augmentation candidates and $K=3$ average augmentations per plan, the total number of training queries to execute is around $1000\cdot3\cdot500 = 1.5$ million.
\sys borrows ideas from factorized IVM~\cite{nikolic2018incremental} and view materialization for factorized queries~\cite{cjt} to aggressively pre-compute and share aggregates between these queries.   
Since each search iteration evaluates every horizontal and vertical candidate, we will first describe optimizations for individual augmentations, and then describe sharing across iterations.

\subsubsection{Horizontal Augmentation}
Given the current augmentation plan $\mathcal{P}$, horizontal augmentation $Q_k^h$ will union the plan with the candidate dataset $D_k^h$.  
We can use IVM to push the aggregation through union~\cite{nikolic2018incremental}.  
\begin{align*}
  Q_k^h = \gamma(\mathcal{P}(T) \cup  D_k^h)
  = \blue{\gamma(\mathcal{P}(T))} \cup  \red{\gamma(D_k^h)}
\end{align*}

\begin{example}
  Suppose we union $\mathbf{R}_1$ and $\mathbf{R}_2$ (\Cref{fig:factorizedex}(a,b)). 
  The gram matrix of the union  is defined as $\gamma(\mathbf{R}_1 \cup \mathbf{R}_2)$ (\Cref{fig:union_gram}),
  which is equivalent to $\gamma(\mathbf{R}_1) \cup \gamma(\mathbf{R}_2)$ (\Cref{fig:factorizedex}(e,f)).

    To fit the linear regression model using the gram matrix  (\Cref{fig:union_gram}), we treat B as feature and C as the target variable. $\theta = [\theta_B, \theta_0]$ is then:
{\small\begin{equation*}
    \theta = (\mathbf{X}^T\mathbf{X})^{-1}\mathbf{X}^TY =
    \begin{bmatrix}
    \sum B^2 & \sum B \\
    \sum B & \sum 1 
    \end{bmatrix}^{-1} \begin{bmatrix}
    \sum BC  \\
    \sum C
    \end{bmatrix} = \begin{bmatrix}
    39 & 11 \\
    11 & 4 
    \end{bmatrix}^{-1} \begin{bmatrix}
    40  \\
    14
    \end{bmatrix}
\end{equation*}}
\end{example}

\noindent The key optimization is to pre-compute \blue{$\gamma(\mathcal{P}(T))$} at the start of the search iteration, and 
\red{$\gamma(D_k^h)$} when data providers upload the dataset. \blue{$\gamma(\mathcal{P}(T))$} is shared across all $M$ candidate horizontal augmentations, and \red{$\gamma(D_k^h)$} is shared across $U$ user requests where $D_k$ is a horizontal candidate. Now, horizontal augmentation simply adds the pre-computed aggregates in near-constant time.

\subsubsection{Vertical Augmentation}
Vertical augmentation is more complex than horizontal augmentation because pushing 
aggregation through the join needs to take the join key $j$ into account (so the join can be evaluated).  
Consider $Q^v_{k}$, which augments the current augmentation plan $\mathcal{P}$ with $D_k^V$ using join key $j$\footnote{ Note for Left Join: When the join keys from vertical augmentations have missing values, performing an inner join will remove tuples and skew training data. Following prior works~\cite{chepurko2020arda} \sys performs a left join between the user and augmentation datasets such that the tuples in the user dataset remain without reducing the cardinality. To impute missing values in the left join result, \sys uses the rules in \Cref{sec:preprocess} and computes annotations for imputed values. \sys does not impute missing values when sending the materialized augmented relation to AutoML, because AutoML systems search for the best imputation method during hyper-parameter optimization~\cite{feurer-neurips15a}.}:
\begin{align*}
  Q^v_k =\gamma(\mathcal{P}(T) \leftouterjoin_{j}  D^v_k) 
  = \gamma(\blue{\gamma_{j}(\mathcal{P}(T))} \leftouterjoin_{j} \red{\gamma_{j}(D^v_k)})
\end{align*}

\begin{example}
  Suppose we have already horizontally augmented $\mathbf{R}_1$ with $\mathbf{R}_2$,
  and want to assess the vertical augmentation $(\mathbf{R}_3,A)$.
  To do so, we want to compute $\gamma((\mathbf{R}_1\cup\mathbf{R}_2)\leftouterjoin_A\mathbf{R}_3)$ (\Cref{fig:join_gram}).
  We can push down the aggregation to derive
  $\gamma\left( (\gamma_A(\mathbf{R}_1)\cup\gamma_A(\mathbf{R}_2))\leftouterjoin_A\gamma_A(\mathbf{R}_3) \right)$ (\Cref{fig:factorizedex}(g,h,i)).
\end{example}

\noindent \blue{$\gamma_{j}(\mathcal{P}(T))$} is shared among all $N$ vertical augmentation candidates with join key $j$. Thus, \sys pre-computes  \blue{$\gamma_{j'}(\mathcal{P}(T))$} for all of its valid join keys $j'$. 
\sys also pre-computes $\red{\gamma_{j'}(D^v_k)}$ for all of its valid join keys, and shares them across all $U$ requests where $D^v_k$ is a vertical candidate. 
Vertical augmentation is now independent of the dataset size and bound by the cardinality of $j$ (typically small).

\subsubsection{Sharing  Between Augmentation Plans} 
Once \sys finds the best vertical augmentation $(D^*, j^*)$ at the end of a search iteration,  
the next plan is defined as $\mathcal{P}'(T) = \mathcal{P}(T) \leftouterjoin_{j^*} \gamma_{j^*}(D^*)$.
At the start of the next iteration, \sys will pre-compute aggregations $\gamma_j(\mathcal{P}'(T))$ for all valid join keys $j$.
Pre-computing $\gamma_j(\mathcal{P}'(T))$ can re-use aggregations from previous plans. 
We illustrate this re-use opportunity using a special case where $j \in S_{D^*} \wedge j \notin S_{\mathcal{P}(T)}$:
\begin{align*}
  \gamma_{j}(\mathcal{P}'(T)) = \gamma_{j}(\mathcal{P}(T) \leftouterjoin_{j^*} D^*) 
  = \gamma_{j}(\red{\gamma_{j^*}(\mathcal{P}(T))} \leftouterjoin_{j^*} \gamma_{j^*,j}(D^*))
\end{align*}

\noindent Here, $\red{\gamma_{j^*}(\mathcal{P}_{i,j}(T_i))}$ has already been computed in a previous search iteration. 
In general, \sys  aggressively pushes aggregations through joins which reduces the join cost and canonicalizes the plans; it then finds subsets of the query plan that is identical to subsets from a previous iteration (as in the example) and re-uses them.

\subsection{Factorized Augmentation Benchmarks}
\label{exp:fac_learning}
Does pre-computation help, or simply add additional overhead?   We now benchmark horizontal and vertical augmentation  to evaluate our pre-computation optimizations (\sys) against naive factorized learning (\textsf{Factorized}).  Specifically, we evaluate 1) what factors influence the performance of \sys?  and 2)  what are the costs of offline pre-computation?

\stitle{Dataset.} We generate a synthetic dataset $T[f_1, f_2, f_3, Y, j]$ with three features, a $Y$ variable, and a join key with a default domain size of $30$.  The feature and $Y$ values are randomly generated. $|T| = 1M$ tuples by default.    We will horizontally and vertically augment $T$.

\subsubsection{Horizontal Augmentation Performance.} 
\begin{figure}
     \includegraphics[width=0.4\textwidth]{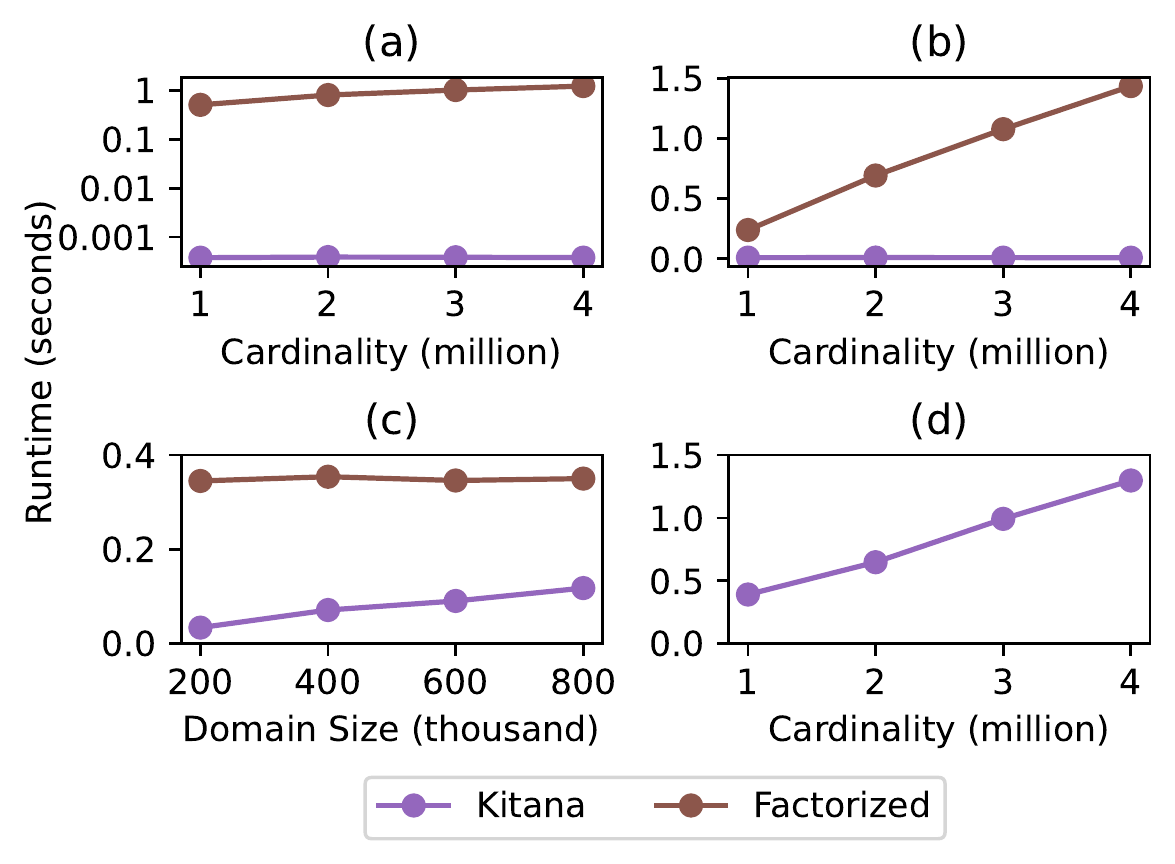}
     \vspace*{-4mm}
     \caption{Factorized Augmentation Benchmarks. (a) Horizontal Augmentation Runtime (log) when varying augmentation dataset cardinality.  
     (b) Vertical Augmentation Runtime when varying user dataset cardinality. 
     (c) Vertical Augmentation Runtime when varying join key domain size. 
     (d) Offline Pre-computation Runtime when varying the cardinality.}
     \label{fig:fac_exp}
     \vspace*{-2mm}
\end{figure}

We create augmentation dataset $D^h$ using the above procedure, vary $|D^h|\in [1M, 4M]$ tuples, and measure the runtime to compute $\gamma(T\cup D^h)$. \Cref{fig:fac_exp}(a) shows that \sys is ${>}3$ orders of magnitude faster than \textsf{Factorized}.  At $|D^h|=4M$, \textsf{Factorized} takes ${\sim}1.2s$ while \sys takes ${\sim}400{\mu}s$.  \textsf{Factorized} grows linearly in  relation cardinality as it needs to compute the aggregates online, while \sys is constant.

\begin{figure}
     \includegraphics[width=0.4\textwidth]{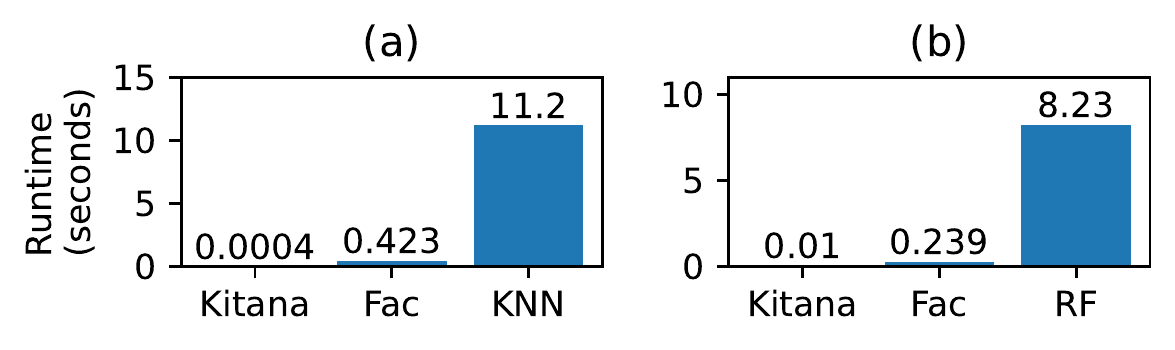}
     \vspace*{-4mm}
     \caption{Factorized Augmentation Compared to models from previous works. \texttt{Fac} is factorized learning without pre-computations.  (a) Horizontal Augmentation Runtime. 
     (b) Vertical Augmentation Runtime. }
     \label{fig:fac_comparison}
\end{figure}

\subsubsection{Vertical Augmentation Performance.}\label{exp:verticalaug} We create augmentation dataset $D^v[j,f]$ containing the join key $j$ and one feature $f$.  The performance of \sys is bounded by the join key domain size and is independent of the cardinality of the query dataset. 
\Cref{fig:fac_exp}(b) varies $|D^v|\in[1M,4M]$ but fixes $|j|=30$, and we see that \sys is constant while \textsf{Factorized} grows linearly.
\Cref{fig:fac_exp}(c)  fixes $|D^v|=1M$ but varies $|j|\in[200k, 400k, 600k, 800k]$.
\sys grows linearly because the aggregates are larger. However, \sys still avoids recomputing aggregates from scratch, and is ultimately upper-bounded by \textsf{Factorized}.  

\subsubsection{Sharing Between Augmentation Plans.} We create three datasets $R_1[A,B,Y], R_2[B,C,f_1], R_3[C,D,f_2]$ with two features, a $Y$ variable, four join keys $A,B,C,D$ with domain size $200k$, and each with $1M$ tuples. We consider original plan $\mathcal{P} = R_1\Join_B R_2$ and new plan $\mathcal{P}' = R_1\Join_B R_2\Join_C R_3$. The task is to compute $\gamma_j(\mathcal{P}')$ for $j \in \{A,B,C,D\}$. Without pre-computation, the four aggregates take $3.13s$ to complete. With aggregates of original plan $\gamma_j(\mathcal{P})$ for $j \in \{A,B,C\}$ pre-computed and re-used, we can compute the aggregates for $\mathcal{P}'$ in $1.75s$, thus $1.8\times$ speedup.

\subsubsection{Offline Pre-computation Overhead}
\sys's pre-computation essentially shifts the cost for \textsf{Factorized} to offline rather than during augmentation search.   \Cref{fig:fac_exp}(d) reports pre-computation time while varying $|D^v|\in[1M,4M]$.   It grows linearly with $|D^v|$, but is only ${\sim}1.3s$ for $4M$ tuples.  This is because the gram-matrix semi-ring decomposes into sum-aggregation operations that are optimized in analytical DBMSes and data science libraries like numpy~\cite{harris2020array}.
Finally, notice that pre-computation is run once per dataset, and can be delegated to the data provider by extending  $\sys.upload(D,l)$.

\subsubsection{Comparison against Other Models.} Although \sys improves considerably over naive factorized learning, prior data augmentation works don't apply factorized learning and are even slower. To highlight the performance disparities, we now compare \sys, factorized learning (\textsf{Fac}), and prior augmentation techniques. 

We consider previous models for both horizontal and vertical augmentations.
For horizontal augmentations, Li et al.~\cite{li2021data} computes novelty by training a 3-NN regressor (\textsf{KNN}) for each augmentation candidate. We test the training time on a horizontal augmentation candidate with $|D^h|= 1M$. \Cref{fig:fac_comparison}(a) shows that \sys is four orders of magnitude faster. For vertical augmentation, ARDA~\cite{chepurko2020arda} joins all tables at once, injects random control features, and trains a random forest to perform feature selection.    We generate $|D^v|$ following \Cref{exp:verticalaug} (without feature injection), and limit ARDA's random forest to a max depth of 3 and 0.1 sampling rate. \Cref{fig:fac_comparison}(b) shows that \sys is three orders of magnitude faster. 

In conclusion, prior works incur unacceptably high search costs that will consume the user's budget.  \sys uses factorized learning and prudent pre-computation to accelerate augmentation search by 2-4 orders of magnitude, at a level that makes data augmentation search for AutoML realistic.

%% file: sections/implementation.tex
\section{\sys Design and Implementation}
\label{sec:imp}

This section describes each component in \sys's architecture (\Cref{fig:service_architecture}). We will present the offline and then online phases.  

\subsection{Offline Phase}

While offline, \sys collects a large volume of datasets for augmentations, preprocesses them, and builds the necessary indexes to serve requests efficiently during the online phase.

\subsubsection{Data Collection.} Enterprise users can make the best of datasets already in the data warehouses by uploading them to \sys with different access labels. They can also upload additional datasets bought from data markets~\cite{awsmarketplace,snowflakedatamarketplace}, or use existing web crawlers~\cite{kausar2013web} to collect public dataset~\cite{nycopen,cms,dataworld} with $RAW$ access labels.

\subsubsection{Data Preprocessing}
\label{sec:preprocess}
\sys preprocesses data corpus offline for efficient online {\it Data Discovery} and {\it Factorized Learning}.

\mypar{Feature Engineering} \sys performs  standard feature engineering~\cite{kuhn2019feature}, including missing values imputation\footnote{In addition to dealing with missing values offline, \sys also applies the missing value imputation methods during online vertical augmentations on tables with join keys containing missing values. This is because \sys uses a left join to avoid reducing the cardinality of the output join,  which can introduce nulls.}, feature transformations (e.g., PCA, polynomial, and interaction features)
and standardization (center and re-scale numeric attributes).
\revise{
We note that  \sys is robust against inadequate missing value imputations because it uses ML performance as the criteria for search. For instance, if vertical augmentation leads to numerous missing values that are hard to impute, the resulting model performance is likely to be compromised, prompting \sys to forgo the augmentation. 
More advanced feature engineering and imputation methods can turn a ``poor'' augmentation into a ``good'' one; we defer the investigation of more advanced methods to future work.
}

\mypar{Data Discovery} Data discovery finds augmentation candidates for a given table.  It does so by computing a profile (e.g., minhash and other statistics) for each table and building an index over them.   \sys uses Aurum~\cite{fernandez2018aurum} by default, which builds a {\it discovery index} offline.   The index finds union-able tables based on syntactic schema matching and value similarity, and join-able tables based on a combination of similarity, containment, and semantic similarity~\cite{fernandez2018seeping} metrics.   \sys can use other data discovery systems~\cite{zhang2020finding,bogatu2020dataset,castelo2021auctus} as well, and they build similar types of indexes offline.

\mypar{Gram Matrix} To support efficient online factorized learning, \sys pre-computes aggregated gram-matrix for each dataset $D$ in the corpus as discussed in \Cref{sec:fac_aug}; the aggregates include $\gamma(D)$ and $\gamma_j(D)$ for each valid join key $j$. Building the gram-matrix has low overhead as shown in \Cref{exp:fac_learning}, and can potentially be offloaded to the data providers during dataset upload.

\mypar{Re-weighting}
For vertical augmentation, one-to-many join results will have a much larger number of tuples and potentially skew the training data distribution. To avoid this, we re-weigh the semi-ring such that, for each join key value, its semi-ring has a count of 1. Given the aggregated gram-matrix semi-ring $(c, s, Q)$ for each join key, we multiply it with $(1/c, 0, 0)$ to $(1, s/c, Q/c)$.

\subsubsection{Handling Updates}

\sys efficiently supports deletes, inserts, and updates of datasets in the corpus. First, the data discovery system we use supports incremental maintenance~\cite{fernandez2018aurum}. Second, we use factorized IVM~\cite{nikolic2018incremental} to update the semi-rings incrementally.

\subsection{Online Phase}
\label{costestimation}

During the online phase, \sys preprocesses the user's base table and checks the request cache in case it can save computation. Then it triggers the augmentation process described in \ref{subsec:algo}, while balancing the budget usage. It uses an AutoML service to find a good model and finally it constructs a response to the user.

\subsubsection{Request Preprocessing}

Given the user's training dataset $T$, \sys computes a profile and uses it to probe the discovery index to find augmentation candidates (L6). By default, \sys splits $T$ into 10-folds for cross-validation, but optionally accepts a user-provided validation set (e.g., in case of train-test imbalance).   At the beginning of each iteration (L4-18), \sys pre-computes $\gamma(\mathcal{P}(T))$ and $\gamma_j(\mathcal{P}(T))$ for all valid join keys to share computations (\Cref{sec:fac_aug}).

\subsubsection{Request Cache}
\label{sec:reqcache}

Caching augmentation plans helps when the training data from two requests match the schema (L2).  We use the most recent plan that improves the performance by ${\ge}\delta$, where $\delta$  ($0.02$ by default) is also used for early stopping ((L15). 
We design a two-level {\it Request Cache}: each schema stores a list of $K$ plans ($K=1$ in the experiments).   The list is ordered and replaced using LRU; a plan is considered used if it improved the model by ${\ge}\delta$. 

\subsubsection{Cost Model}
\label{sec:costmodel}
\sys shortcuts to AutoML once there is not ``enough time'' left (L16) predicted by a cost model. 
The cost model $f(n,m)\to \mathbb{R}^+$ maps the shape $n\times m$ of the augmented training dataset $T'$ as input, and outputs the time to train user-requested model $K$ times ($K=5$ by default).   
The function should over-predict to ensure \sys is not worse than AutoML without augmentation search, and is typically tuned to specific hardware and model types.

Since \sys uses Auto-sklearn~\cite{feurer-neurips15a} as its default AutoML library, we use 
scitime~\cite{scitime} to construct the cost model. Scitime is an open-sourced project for training time estimation for Sklearn algorithms given training data shape; we found it accurate in our experimental evaluation.
Scitime can be customized to other AutoML systems (e.g., FLAML~\cite{wang2021flaml}) by executing them on randomly generated data of varying sizes, collecting training time, and training a model (defaulting to random forests) to predict training time.
ML cost estimation is an active area of research~\cite{popescu2013predict,wang2015performance}, and we expect cost estimators to improve over time. 

\subsubsection{Prediction API Construction}
\label{sec:apiconstruct}
\sys constructs and exposes a prediction REST API if the request $R$ includes $API$ in the return labels.  
The API takes as input a dataset $T'$ with the same schema as $T$, and outputs the prediction values for each record. Internally, it applies all vertical augmentations from $\mathcal{P^*}$ to $T'$, and then uses the final model $\theta$ to perform the predictions.  If \sys internally uses a cloud AutoML service  rather than Auto-sklearn, then \sys will call the service's own prediction API.  Note that this will incur monetary charges that the user is responsible for.

%% file: sections/evaluation.tex
\input{sections/tautoml}

\section{Evaluation}
\label{sec:evaluation}

Our evaluation strives to understand three main questions:
\textbf{Q1}: Can \sys improve task performance beyond real-world AutoML services?
\textbf{Q2}: How does \sys adapt to varying budgets and the percentage of useful datasets?
\textbf{Q3}: how sensitive is \sys performance to different components and their configurations?

\subsection{Setup}
\label{exp:setup}
\mypar{Datasets} We collected 518 datasets from NYC open data~\cite{nycopen} (364 datasets, 3.1 GB) and CMS Data~\cite{cms} (154 datasets, 7.4 GB). To create user requests, we adopt a leave-one-out strategy: 1 dataset is chosen for AutoML, and the rest 517 datasets are searched for augmentation. For each requst, the dataset  is split into 80\% training data and 20\% testing data.

\stitle{AutoML:} 
\begin{myitemize}
\item \revise{\textbf{SK}: Auto-sklearn~\cite{feurer-neurips15a} is an open-source library that performs hyperparameter optimization over a broad range of ML models including multilayer perceptron, KNN, SVM, tree-based models (decision tree, gradient boosting and random forests).}

\item \revise{\textbf{FML}: FLAML~\cite{wang2021flaml} is also an open-source AutoML library, but it  uses mostly tree-based models. }

\item \textbf{Vertex AI}~\cite{Vertexai} is a cloud-based AutoML service by Google. It currently has a minimum time budget of 1 hour and doesn't enforce the budget. Therefore, we run Vertex AI until completion.
\end{myitemize}

\stitle{Baselines:} 
\begin{myitemize}
\item \textsf{AutoML-Only} directly sends the request to an AutoML service (\textbf{SK}, \textbf{FML}, or \textbf{Vertex AI}) without augmentations. 

\item \sys (\textbf{K}) applies our full suite of optimizations to speed up the augmentation search, uses the cost model balances the budget split, and sends the augmented data to AutoML. 
\revise{\textbf{K} uses the cost model in \Cref{sec:costmodel} to balance between augmentation and AutoML. For Vertex AI, predicting the cost is challenging. We, therefore, allocate a fixed  budget of $5min$ for \textbf{K},  which, while small compared to the ML cost (${>}1hour$), shows sufficient for \textbf{K} to find predictive augmentations in our experiments.
}

\item \textsf{Factorized}  (\textbf{Fac}) is similar to \sys, but doesn't pre-compute sketches.  It computes the aggregates online and fully re-trains {\it factorized proxy models} to assess candidate augmentations.

\item \revise{\textbf{ARDA}~\cite{chepurko2020arda} is the SOTA data search algorithm for ML,  which joins candidate augmentations (with pre-aggregations to avoid many-to-many joins), injects random features, and trains models (random forests and sparse regression) over the join to assess features. We run ARDA with its default setting (20\% random injected
features, $10$ rounds of injection, random forests with 3 max depth, 10\% sampling rate, and $100$ trees).}

\end{myitemize}

\mypar{Setup} All experiments run in a single thread on a GCP n1-standard-16 VM, running Debian 10, Xeon 2.20GHz CPU, and 60GB RAM. We implemented \sys in Python 3.  To compute monetary costs, we report computation costs because they dwarf storage costs. For instance, Google Cloud storage costs \$0.046/GB/month, while running an n1-standard-16 instance costs about $\$388.36$/month.

\subsection{Q1: Does \sys use Data Effectively?}
\label{sec:data_service_value_exp}

\revise{We first evaluate requests using six representative datasets (regression/classification tasks~\footnote{\sys supports binary classification by treating it as a thresholded regression~\cite{peng2020discriminative} where the two labels are represented as numerical values 0 and 1.},  varying sizes) for detailed insights.  Then, we scale up to all 518 datasets via a leave-one-out strategy. All requests are run 10 times.}

\subsubsection{Exploratory  Study}
\label{exp:exploratory}
We first conduct an exploratory comparison of \sys with existing AutoML services and data augmentation search algorithms on price and model performance, and show that \sys can 1) reach the same or greater model accuracy for considerably lower cost,   2) reach far higher model accuracy for the same cost, and 3) often times reach a higher accuracy at a lower cost.

\stitle{Setup.} \revise{We construct requests (regression, classification) using six datasets with varying sizes (${\leq}1MB$, $1{-}10MB$, ${>}10MB$). Model performance measures $R2$ for regression and accuracy for classification\footnote{Let $y_i$ represent the target variable for the $i^{th}$ tuple, with $\hat{y}_i$ as its predicted value. Let $n= \sum_i 1$ denote the total number of tuples and $\bar{y}=\sum_i y_i/n$ be the average target variable. For regression, $y_i$ and $\hat{y}_i$ are numerical, while they are 0/1 for classification. Then $R^2=1-\frac{\sum_i(y_i-\hat{y}i)^2}{\sum_i(y_i-\bar{y} )^2}$, and $accuracy=\sum_i\mathbf{1}_{\{y_i=\hat{y}i\}}/n$, where $\mathbf{1}$ is the indicator function equal to 1 when $y_i=\hat{y}_i$ and 0 otherwise.}.
We evaluate Auto-sklearn (\textbf{SK}), FLAML (\textbf{FML}) and \textbf{Vertex AI}~\cite{Vertexai}, with or without data augmentation search (using \textbf{K}, \textbf{Fac} or \textbf{ARDA}). Since \textbf{SK} is not hosted, we evaluate \textbf{SK} and \textbf{K+}/\textbf{Fac+}\textbf{SK}  on a GCP n1-standard-16 VM and fix the time budget of $10$min  ($\$0.13)$; we find that $10$min  is sufficient for model convergence. \textbf{ARDA} exceeds $10$min, and we cap its runtime at $2$ hours.  \textbf{Vertex AI} is set to its minimum budget of 1 hour, though it exceeds the requested budget.  We report  results over 10 runs.}

\stitle{Results.}
\revise{\Cref{exp:vertexai} summarizes the min/max model performance (Score) from 10 runs, runtime in minutes (Time), and the cost in USD (Cost); we highlight the highest model performance in \red{\textbf{red}}. 
We find that the model performance was consistent across all 10 runs (min/max difference ${\leq}0.01$).
\textbf{SK} and \textbf{FML} exhibit similar performance. \textbf{Vertex AI} also has comparable model performance but takes an order of magnitude longer to run, so we did not report it.}

\indent \revise{Adding data augmentation (\textbf{K+}) improves the model quality by a large margin (e.g., $0.16{\to}0.91$ for $D_2$~\cite{gender}) for most user requests.  However, for $D_6$\cite{y2hd-n93e}, there aren't many augmentations to greatly improve model performance.
Overall, \sys provides an opportunity to balance data augmentation and AutoML.  
Compared to \textbf{K}, \textbf{Fac} takes longer to find augmentations as it doesn't pre-compute sketches. This results in less time for dataset search, which explains its lower performance.
\textbf{ARDA} is comparable to \textbf{K} but much slower and costlier due to join materialization and expensive model training---\textbf{ARDA} times out for $D_3$ and $D_6$.}

\begin{figure}
     \includegraphics[width=0.45\textwidth]{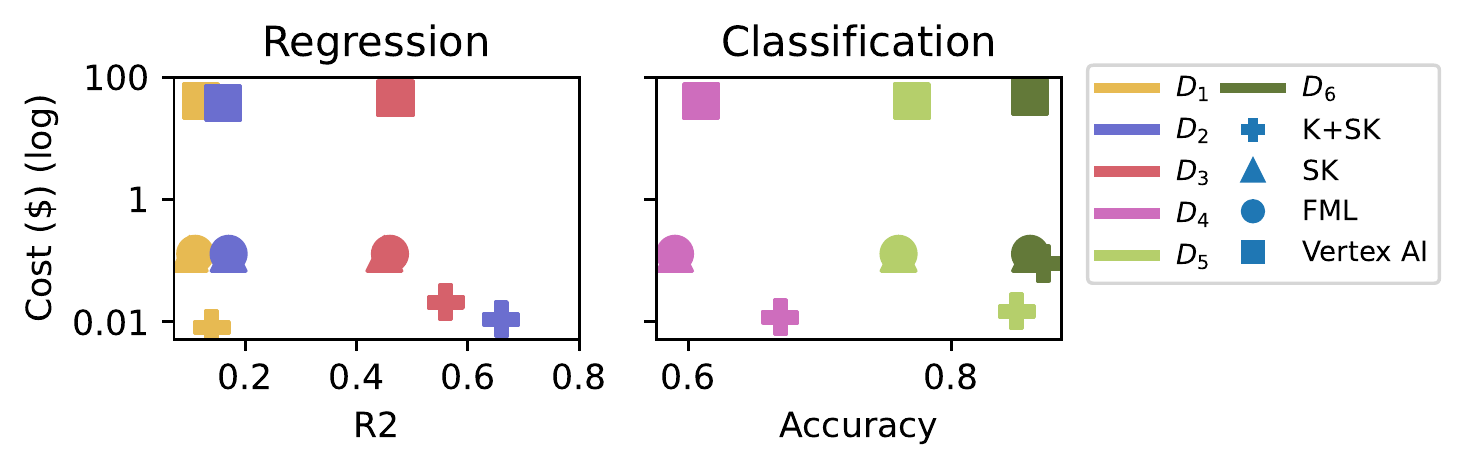}
     \vspace*{-4mm}
	 \caption{\revise{Cost for \sys to reach or exceed the model quality of Auto-sklearn and Vertex AI (shapes) for different user requests (color).  \sys trains higher quality models ${\sim}10\times$ (${>}100\times$) cheaper than AutoSklean (Vertex AI). }}
     \label{fig:marketcost}
     \vspace*{-3mm}
\end{figure}

\indent \revise{\Cref{fig:marketcost} reports the cost for \sys to achieve the same or higher model quality as \textbf{SK}, \textbf{FML} and \textbf{Vertex AI} (lower right is better).  In every case, \sys is better. For $D_2$, \sys reduces the cost from $\$0.13$ (\textbf{SK}) or $\$38.6$ (\textbf{Vertex AI}) to merely $\$0.01$, while improving the $R2$ from $0.17$ to $0.66$.}

\begin{figure}
 \centering
 \includegraphics[width=0.45\textwidth]{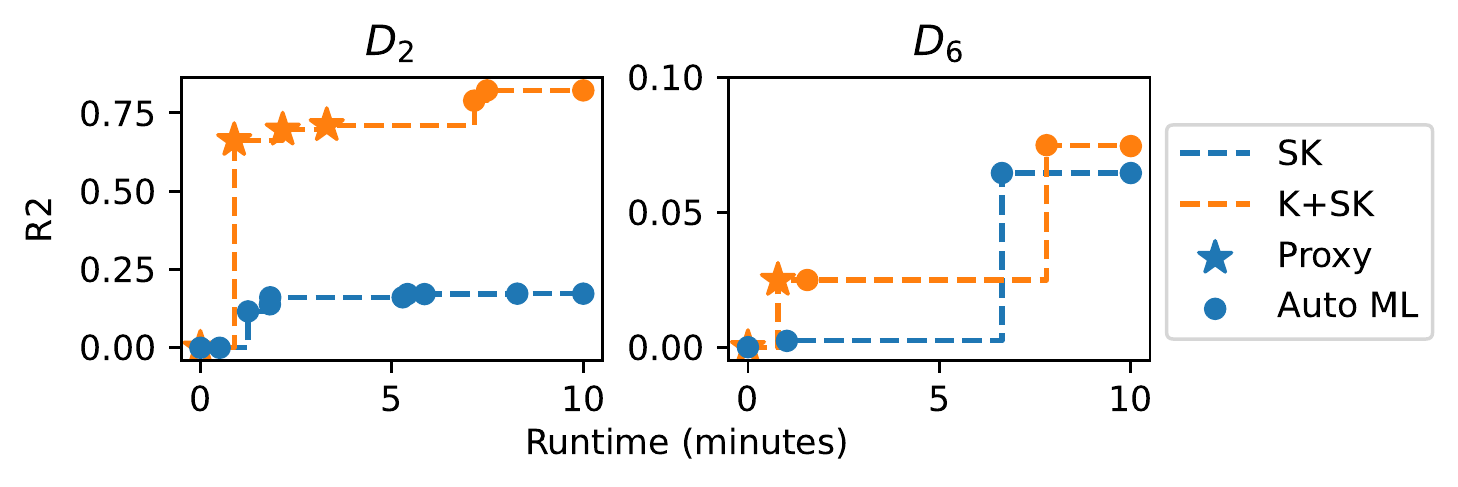}
\vspace*{-5mm}
 \caption{\revise{Baseline performance (testing R2) over time. $\bigstar$ is for proxy model  and $\bullet$ is for AutoML model performance.}}
 \label{exp:service_value}
\end{figure}

\subsubsection{In-depth Study}
\label{exp:dep}
\revise{We now study requests $D_2$ and $D_6$, which respectively showed the most and least improvement from dataset search. $D_2$ predicts the ELA test scores across schools in 2013-2016, while $D_6$ predicts the ownership type of nursing homes. 
Since \textbf{SK} incrementally adds models to an ensemble, we are able to report its $R2$ over time. For \sys, we also report the proxy model in each augmentation search iteration. }

\indent \revise{\Cref{exp:service_value} plots model performance (y-axis) by runtime (x-axis) for \textbf{SK} with and without \textbf{K} (color).  We report the proxy (star) and AutoML (circle) models.  
For $D_2$, \sys reaches an $R2$ of ${>}0.6$ in less than 1 minute, whereas AutoML-only never exceeds $0.2$ even after 10 minutes. This highlights the importance of high-quality features over pure compute~\cite{sambasivan2021everyone,chepurko2020arda}. Even though $D_6$ is $35MB$, \sys can still find a useful augmentation within a minute because its cost is proportional to the join key's domain size and not data size (\Cref{exp:fac_learning}).  \sys quickly switches to AutoML as no more datasets improved the proxy's $R2$ by $\delta{=}0.02$ (L15). }

\indent  \revise{We found that the datasets in the final augmentation plans were logically relevant to the requests.  For $D_2$, the sequence of datasets are: school quality report~\cite{cbfr-z7aj} (Student Attendance Rate, Percent of English Language Learners), 
neighborhood demographic distributions~\cite{iuvu-z276},
and historical math test scores~\cite{b9uf-7skp} from 2006-2012.   These are all relevant to ELA test scores.
\sys augmented $D_6$ with provider information~\cite{provider} (number of beds, quality measure scores), but we could not find other related datasets in our corpus.}

\subsubsection{Comprehensive Study}
\label{exp:com}
\revise{To test the generalizability of our previous findings, we compare \textbf{SK}, \textbf{K+SK}, and \textbf{Fac+SK} using all 518 datasets; for each dataset in the corpus, we remove it from the corpus and use it as the request.  We set the time budget to $10$min, and report the median of 10 runs.}

\begin{figure}
\begin{subfigure}[b]{0.24\textwidth}
         \centering
         \includegraphics[width=\textwidth]{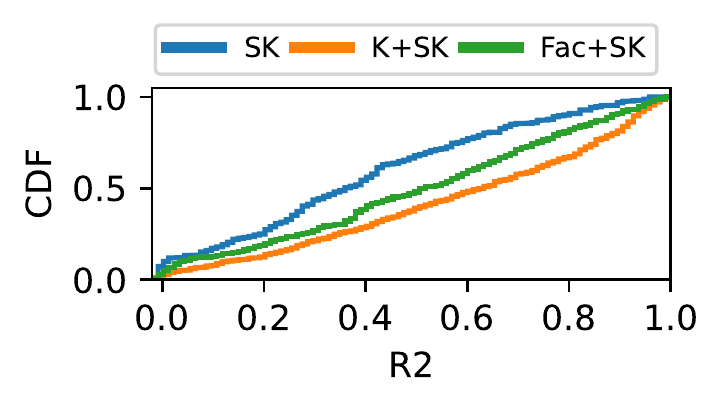}
         \vspace*{-6mm}
         \caption{\revise{Cumulative distribution function (CDF) of model performance.}}
         \label{fig:cdf}
         \vspace*{-2mm}
\end{subfigure}
\begin{subfigure}[b]{0.23\textwidth}
         \centering
         \includegraphics[width=\textwidth]{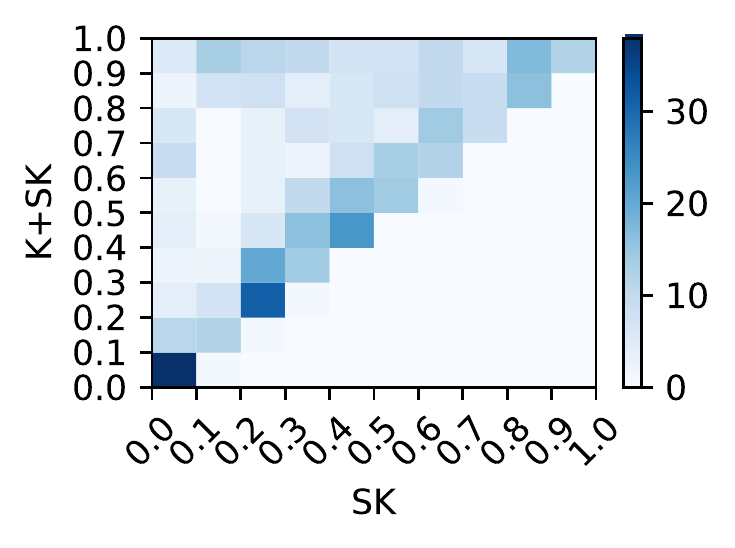}
         \vspace*{-6mm}
         \caption{\revise{Heatmap of the model performance between \textbf{K+SK} and \textbf{K}.}}
         \label{fig:heatmap}
         \vspace*{-2mm}
\end{subfigure}
    \caption{\revise{Model performance  over 518 datasets using leave-one-out strategy. (a). Both \textbf{K+SK} and \textbf{K+FML} show comparable performance and demonstrate significant improvements over \textbf{K}. \textbf{Fac} falls in the middle, as it requires more time for searching. (b).  \textbf{K+SK} outperforms \textbf{K} (above diagonal) on a range of datasets, with a long tail of significant improvements (upper left corner) up to $0.85$.}}
    \label{exp:compre}
\end{figure}

\indent \revise{\Cref{fig:cdf} shows the cumulative density function for model performance (lower curve is better). \textbf{K+SK} and \textbf{Fac+SK} respectively improve the median R2 by $0.25$ and $0.17$ over \textbf{SK} alone. 
\Cref{fig:heatmap} displays the heatmap of model performance across \textbf{SK} (x) and \textbf{K+SK} (y). \textbf{K} never degrades model performance (all points are above the diagonal).   Although some datasets report low model performance and negligible improvements from dataset search (bottom left corner), there is a long tail of large model improvements (top left corner) of up to $0.85$.}

\subsection{Q2: \sys Adaptability}
\label{exp:stability}

Our previous experiments used a generous time budget and a corpus that contains good augmentation candidates.  We now study how \sys adapts under varying time budgets, and when there are no good augmentation opportunities.

\subsubsection{Varying Budget} 
\label{exp:budget}
\revise{We repeat the experiment from \Cref{sec:data_service_value_exp} but vary the time budget of each request to $\{2,4,6,8,10\}$ minutes.  We run each request 10 times and report the median performance.}

\begin{figure}

    \begin{subfigure}[b]{0.5\textwidth}
         \centering
         \includegraphics[width=0.9\textwidth]{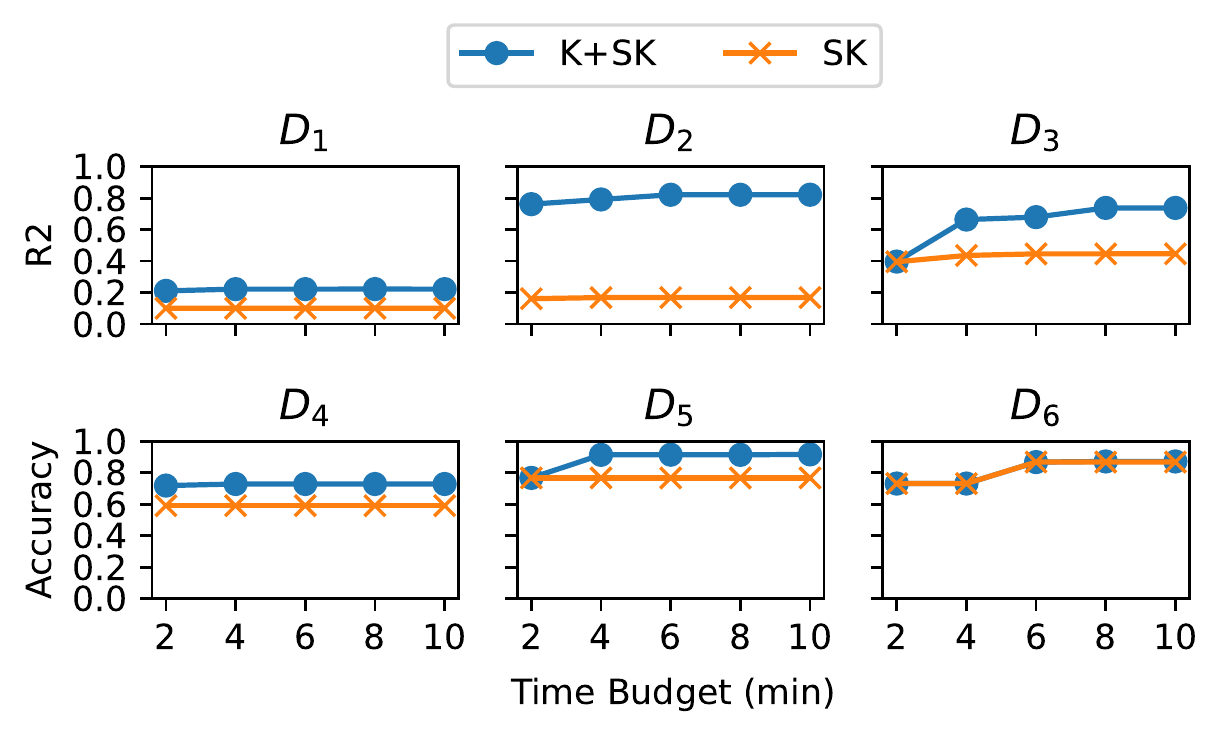}
         \vspace*{-2mm}
         \caption{\revise{\sys performance when varying time budgets. \sys adapts to lower budgets by choosing cheaper but effective augmentations. }}
         \label{fig:marketbudget}
     \end{subfigure}

     \begin{subfigure}[b]{0.5\textwidth}
         \centering
         \includegraphics[width=0.9\textwidth]{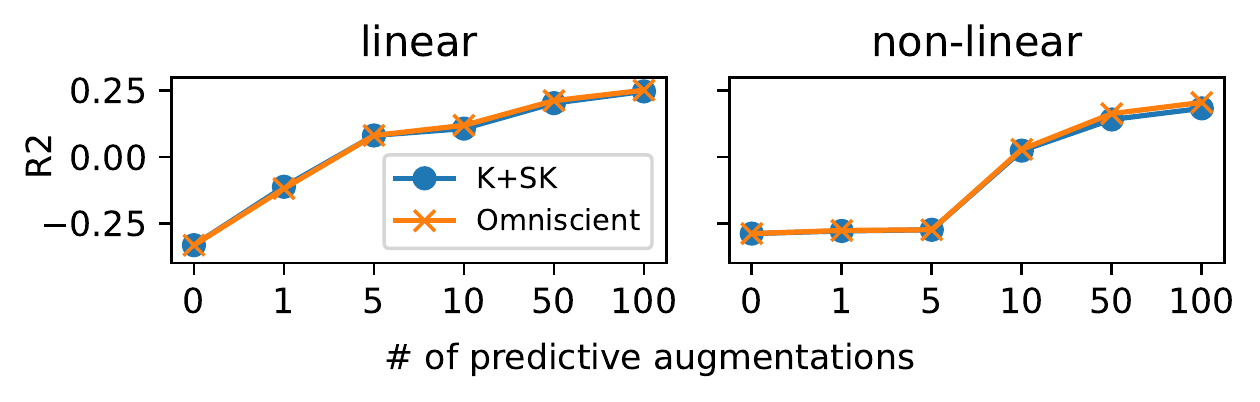}
         \vspace*{-3mm}
         \caption{\revise{\sys performance on synthetic data corpus when varying the number of predictive augmentations. \sys is able to identify these augmentations and achieve similar performance to the omnious search (upper bound) due to the effectiveness of proxy  models.}}
         \label{fig:vary_predictive_aug}
     \end{subfigure}
    \caption{\revise{\sys adaptability experiments.}}
\end{figure}

\indent \revise{\Cref{fig:marketbudget} shows that \sys adapts to shorter budgets by finding augmentations that require less AutoML training time. Small datasets (${<}1MB$) require less training time and \sys improves model quality even when limited to $2$min. For larger datasets, \sys uses the cost model to allocate enough time for AutoML, and improves over \textbf{SK} considerably when there is enough time to accomodate augmentations.  $D_6$ is an exception because the corpus does not contain useful augmentations.   In all cases, \sys never performed worse than \textbf{SK}.
}

\subsubsection{Vary Predictive Augmentations.} 
\label{exp:pred}
\revise{How does \sys perform when the predictive augmentations are limited or abundant?
Can \sys effectively identify all the predictive augmentations, compared to an omniscient search engine that knows all predictive augmentations and doesn't have time constraints? 
Addressing these is challenging for real-world datasets as the ground truth of predictive augmentations is unknown. To overcome this, we employ synthetic experiments that create predictive augmentations.

\stitle{Setup.} For user request, we create a dataset $R[y, J_1, ..., J_{10}]$ of $1M$ rows, where $J_i$ join key of  random integers in $[1, 10K]$. We create $10$ ground truth feature tables $F_i[J_i, f_i]$ of $10K$ rows, where $f_i$ are random floats in $[0,1]$. For $y$, we consider either a linear ($y=\sum_{i=1}^{10}f_i$) or a non-linear relationship ($y=\sum_{i=1}^{10}f_i^2$) with the features.}

\indent \revise{We create both horizontal and vertical augmentations. We introduce a train-testing imbalance, where the test and validation datasets are uniform samples of $10K$ rows from $R$, whereas the training dataset is a non-uniform partition of $R$: we make the first feature $f_1$ public, and divide $R$ into $11$ partitions evenly based on $f_1$.  The first partition is used as training, and the rest are horizontal augmentations.
For the vertical augmentation, we create noisy versions of $f_2,...,f_9$ to simulate the real-world corpus: for each feature, we create $10$ predictive augmentations $A_i[J_i, c_i]$, where $c_i$ varies in correlation with $f_i$~\cite{kaiser1962sample}. Specifically, the correlation coefficient $\phi$ is drawn from the inverse exponential distribution $\text{min}(1, 1/\text{Exp}(10))$, and $c_i$ is the weighted average between $f_i$ and a random variable, weighted by $\phi$. There are $100$ predictive augmentations in total ($10$ horizontal and $90$ vertical).}

\indent \revise{We build a data corpus of $100$ datasets, and vary the number of predictive augmentations $\{0,1,5,10,50,100\}$ in them
by choosing random samples from the $100$ predictive augmentations; the rest datasets are union-able or join-able relations but populated with random numbers. 
We run \textbf{K+SK} with a $10$min  budget, and an Omniscient search procedure that joins all predictive augmentations and runs AutoML until convergence without time constraint. }

\revise{\stitle{Result.} \Cref{fig:vary_predictive_aug} shows the results. The baseline training data is unbalanced, so the starting points ($R2$) for both are negative.  However, with more predictive augmentations, \sys can find them and achieve near-identical performance to Omniscient search ($R2$ difference $\leq0.01$), for linear and non-linear relationships. }

\begin{table}
\begin{center}
\begin{tabular}{  c  rrr } 
 & \textbf{Time} & \textbf{Training R2} & \textbf{Testing R2}\\ 
  \textbf{\sys} & 0.01s & 0.995 & 0.994 \\
  \textbf{Novelty} & 9.72s &  0.773 & -0.232 \\
\end{tabular}
\end{center}
\caption{Model performance for horizontal augmentation. }
\label{exp:horizontal_metric}
\vspace*{-4mm}
\end{table}

\subsection{Q3: Component Sensitivity}

We now evaluate \sys when restricted to horizontal augmentation only against recent data augmentation work by Li et al.~\cite{li2021data}, and also evaluate the effects of request caching across user requests.  

\subsubsection{Horizontal Augmentation Comparisons}
\label{sec:lidata}
Recent works, such as the \textsf{Acquisition} algorithm by Li et al.~\cite{li2021data}, also study the problem of horizontal augmentation.  At a high level, given a corpus of datasets, \textsf{Acquisition} seeks sample ``novel'' records from datasets, based on a custom ``Novelty'' metric.   Novelty is estimated by unioning subsets of the user and  augmentation candidates, and fitting a 3-NN classifier to predict which dataset a given record is from.  \revise{The intuition is that "novel" data can be more easily classified since it exhibits distinct characteristics.}

The key limitation of the novelty measure is that it is oblivious to the actual prediction task, as defined by the test distribution.  If the data corpus offers union-compatible data that is irrelevant to the test distribution (common for the data lake in the wild), this data will have high novelty simply if they are dissimilar to the training data.
In contrast, \sys directly evaluates augmentations with respect to the training data through cross-validation, which is expected to be representative of the test data.

\stitle{Setup.}  We compare \sys with a variation that replaces Lines 13-14 in Algorithm 1 to rank augmentation candidates using \textsf{Novelty}.  This also means that augmentation search does not benefit from factorized learning.  
We measure the time to search for the top one horizontal augmentation candidates, and the model accuracy after running AutoML for $2$ minutes over the augmented dataset.  
Following Li et al.~\cite{li2021data}, we use the \textbf{RoadNet}~\cite{roadnet} mapping dataset (434,874 tuples) with schema \texttt{(lat, lon, altitude)}, and partition it along \texttt{lat} and \texttt{lon} into a $8\times 8$ grid.     Each partition is a candidate horizontal augmentation.
The regression task uses \texttt{lat}, \texttt{lon} to predict \texttt{altitude}, and 
the user's training $T$ and testing $V$ datasets are both 0.5\% samples from partition 1.    In this way, most horizontal augmentations are dissimilar but irrelevant.  

\textsf{Acquisition} uses a reinforcement learning framework to estimate a dataset's novelty because there is a cost to sample from each augmentation candidate.  In our case, we directly evaluate the true novelty in order to study the upper bound of their approach.

\stitle{Results.}
\Cref{exp:horizontal_metric} reports that \sys is orders of magnitude faster than  \textsf{Novelty} because the latter trains a classification model for each augmentation but does not benefit from factorized learning.  Although both approaches have a high R2 on the training data, \textsf{Novelty} prefers augmentations {\it dissimilar to} partition 1, which skews the augmented training data and leads to a low testing R2 score.

\subsubsection{Request Cache.}

Our experiments so far focused on a single user request, however an AutoML service is expected to
serve many requests over time, from the same or different users that perform the same or different tasks.  How much does 
the request take advantage of the similarity between multiple user requests?  

\stitle{Setup.} We create a synthetic benchmark consisting of 20 users and 300 candidate vertical augmentations per user, for a total of 6000 augmentation candidates.  Each user dataset requires 2 augmentations to train a perfect proxy model (R2=1). 
To test the case of failed cached augmentation plans, we group the users into 10 pairs and assign each pair a unique schema.  Thus, each pair of users ($A$, $B$) will match each other's cached augmentation plans, but $A$'s plan will not improve $B$'s model and vice versa.   

We generate sequences of 50 user requests, where the $i^{th}$ user is drawn from a Zipfian distribution over the 20 users.  We vary the parameter $\alpha\in[0,7]$; $0$ is uniform, while $7$ is heavily skewed.  To pressure test the request cache, we limit the cache size to accommodate 5 schemas and 1 plan per schema.

\begin{figure}
     \begin{subfigure}[b]{0.17\textwidth}
         \centering
         \includegraphics[width=\textwidth]{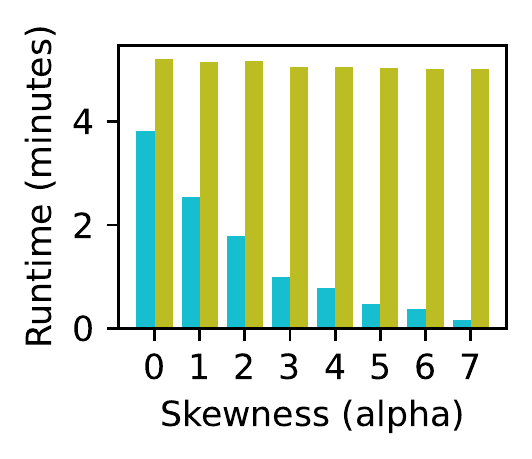}
         \vspace*{-8mm}
         \caption{Sequence Runtime.}
         \label{fig:sequence_runtime}
     \end{subfigure}
     \begin{subfigure}[b]{0.3\textwidth}
         \centering
         \includegraphics[width=\textwidth]{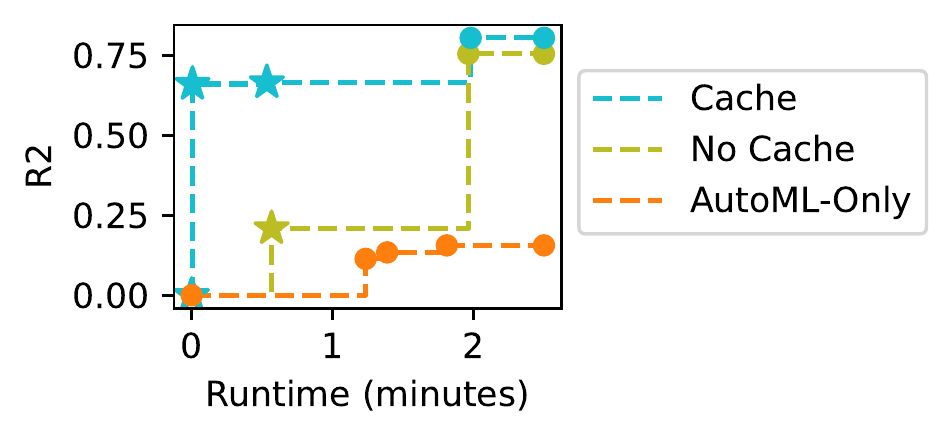}
         \vspace*{-8mm}
         \caption{Gender request with cache.}
         \label{fig:marketcache}
     \end{subfigure}
    
    \begin{subfigure}[b]{0.45\textwidth}
         \centering
         \includegraphics[width=\textwidth]{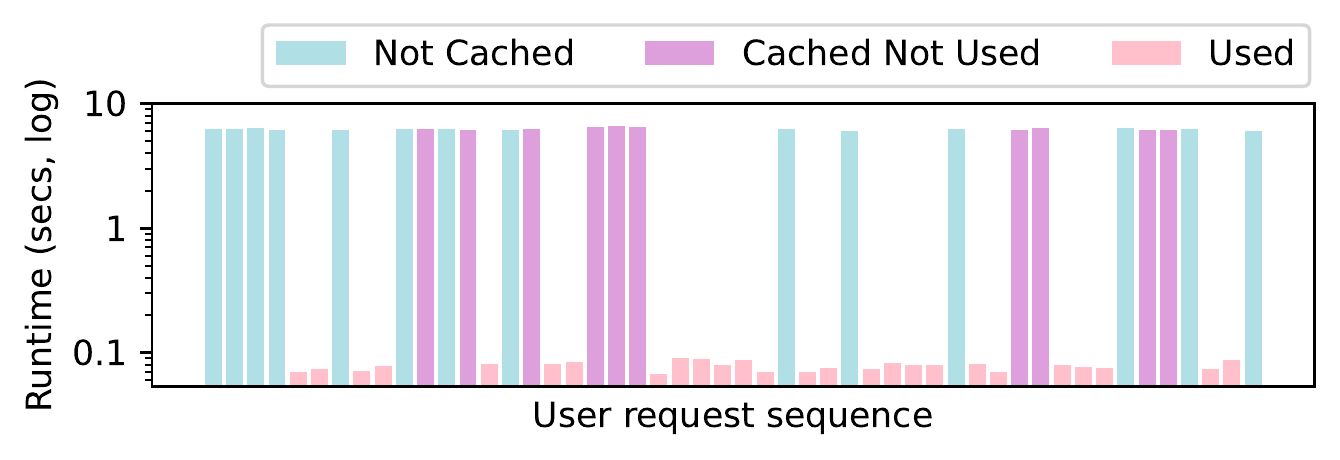}
         \vspace*{-8mm}
         \caption{Cache performance (log) zoom in for alpha = 1.}
         \label{fig:sequence_zoomin}
         \vspace*{-3mm}
     \end{subfigure}
    \hfill
     \caption{Request Cache Experiment Result. The runtime of a sequence of user requests has been reduced because of the cache. The reduction in run-time can improve final model performance as \sys has more budgets.}
\end{figure}

\stitle{Results.} 
\Cref{fig:sequence_runtime} reports the runtime for all 50 requests as the Zipfian skew increases.  Compared with no caching, the cache reduces the runtime by $26.8\%$ given a uniform distribution ($\alpha=0$) and by $96.7\%$ when the requests are nearly all the same ($\alpha = 7$).

\Cref{fig:sequence_zoomin} examines the per-request runtimes when $\alpha = 1$.
We see that cache hits reduce the request runtime by over $100\times$, and also that evaluating unhelpful cached plans (purple) incurs negligible overhead (${\sim}1\%$) as compared to a cache miss (blue).

A cache hit both reduces the augmentation search time, and gives AutoML more time to find a better model.  To see this, we first cache the augmentation plan generated by the \textbf{Gender} user request from \Cref{exp:stability} that has a 5-minute budget.  
We then generate a similar request with a resampled training dataset and a reduced budget of 2.5 minutes. 
\Cref{fig:marketcache} shows that the cached plan immediately increases the proxy model's quality to an R2 of ${\sim}0.7$.  This leads to a better final augmentation plan, and ultimately a higher quality model as compared to a no-cache setting (${\sim}0.75\to{\sim}0.8$).

%% file: sections/tautoml.tex
\begin{table*}[h!]
\small
\setlength{\tabcolsep}{1pt}
\begin{tabular}{ccc|ccccc|cccccc} 
\multicolumn{3}{c}{} & \multicolumn{5}{c}{\textsf{\bf Meet time budget ($10$ minutes)}} & \multicolumn{6}{c}{\textsf{\bf Exceed time budget}} \\
\multicolumn{3}{c}{} & \textsf{\bf SK} & \textsf{\bf FML} & \textsf{\bf Fac + SK} & \textsf{\bf K + SK} & \textsf{\bf K + FML} & \multicolumn{3}{c}{\textsf{\bf ARDA + SK}} & \multicolumn{3}{c}{\textsf{\bf K + Vertex AI}} \\
\textbf{Data} & \textbf{Task} & \textbf{Size} & \textbf{Score} &\textbf{Score} & \textbf{Score} & \textbf{Score} & \textbf{Score} & \textbf{Score} & \textbf{Time} & \textbf{Cost} &\textbf{Score} & \textbf{Time} & \textbf{Cost} \\
$D_1$~\cite{vacc} & R & 0.8MB & [0.10,0.10] & [0.11,0.11] & \red{\textbf{[0.22,0.22]}} & \red{\textbf{[0.22,0.22]}} & \red{\textbf{[0.22,0.22]}} &\red{\textbf{[0.22,0.22] }}& [23,23] & [0.30,0.30] & \red{\textbf{[0.22,0.22]}} & [115,124] & [40.73,40.92]\\
$D_2$~\cite{gender} & R & 4.1MB & [0.17,0.17] &[0.17,0.17] & [0.55,0.55] & [0.82,0.82] & [0.82,0.82] & [0.81,0.82] & [59,61] & [0.77.0.80] & \red{\textbf{[0.91,0.91]}} & [116,131] & [41.09,46.4]\\
$D_3$~\cite{nb39-jx2v} & R & 33MB & [0.45,0.45] & [0.46,0.46] & [0.68,0.68] & \red{\textbf{[0.74,0.74]}} & [0.73,0.74] & & OOT & & [0.70,0.71] & [125,150] & [44.28,53.13]\\
$D_4$~\cite{uk6z-hem9} & C & 0.7MB & [0.60,0.60] & [0.60,0.60] & [0.71,0.71] & \red{\textbf{[0.73,0.73]}} & \red{\textbf{[0.73,0.73]}} & [0.72,0.73] & [22,22] & [0.29,0.29] & [0.70,0.70] & [118,127] & [41.80,44.98]\\
$D_5$~\cite{econ} & C & 10MB & [0.77,0.77] & [0.77,0.77] & [0.89,0.89] & [0.91,0.92] & [0.94,0.94] & [0.92,0.92]& [64,66] & [0.83,0.86] & \red{\textbf{[0.98,0.98]}} & [117,133] & [41.44,47.11]\\
$D_6$~\cite{y2hd-n93e} & C & 35MB & [0.85,0.85] & [0.85,0.85] & [0.85,0.85] & [0.87,0.87] & [0.87,0.87] & & OOT & & \red{\textbf{[0.87,0.88]}} & [131,148] & [46.40,52.42]\\
\end{tabular}
\caption{\revise{AutoML performance across different baselines. 
The performance score ($R2$ for regression and accuracy for classification) is measured in 10 runs, and we report the [minimum, maximum].   The baselines are categorized based on whether they meet or exceed the time budget, and  the best performance is highlighted in red. The 
 (\textbf{K+}/\textbf{Fac+}) \textbf{SK} and \textbf{FML} are allocated a budget of 10 minutes  (cost \$0.13). \textsf{K+Vertex AI}'s budget is set to 1 hour (the minimum allowed) but it exceeds the budget. Due to the protracted time taken by \textbf{ARDA}, we extended its budget to 2 hours, though it still runs out of time (OOT) for larger datasets. The final results show that \sys (\textbf{K+SK/FML/Vertex AI}) outperforms the other baselines with high performance at lower costs.}}
\label{exp:vertexai}
\vspace*{-3mm}
\end{table*}

%% file: sections/relatedwork.tex
\section{Related Work}
\label{sec:relatedwork}

We now distinguish \sys from related work.  

\stitle{AutoML:}
\revise{AutoML services allow non-ML experts to train high-quality models. AutoML is time-critical as it charges hourly (e.g., Vertex AI charges $\$21.252$/hr for tabular datasets). 
Existing AutoML services~\cite{Vertexai,sagemaker, MSautoml, h2o} are {\it model-centric}: 
the training dataset is {\it fixed} and all work is focused on transforming the training dataset (e.g., PCA, polynomial, and interaction features) and model search. As a result, the final model is only as good as the quality of the training dataset.
\sys provides an API-compatible interface to existing AutoML, but is {\it data-centric} that further searches for {\it new} tabular datasets to augment the training data with new features and/or examples. Our experiments show that \sys is orders of magnitude more cost-effective than existing AutoML alone, and achieve almost the same result when no helpful data augmentation is present.}

\stitle{Model Augmentation:}
Recent works propose horizontal or vertical augmentation techniques to improve ML models. For instance, Li et al.~\cite{li2021data} focuses on horizontal augmentation.  They sample records from a set of relations based on a ``Novelty'' measure, assuming that record diversity will improve model accuracy, which may not hold for the data lake in the wild. Our experiments show that the model accuracy can decrease after augmentation.

Other works such as ARDA~\cite{chepurko2020arda} apply specialized feature selection techniques that require iterative training of complex models (e.g., random forest). These techniques are ill-suitable for time-critical tasks such as AutoML: they take orders of magnitudes more time than \sys, and even at such high costs, they empirically don't show better results than \sys (\Cref{fig:moneyfig}).

\stitle{Data Discovery:} Nowadays data warehouses and data lakes leverage data discovery~\cite{fernandez2018aurum} or catalogs~\cite{cataloggoogle,catalogms,collibra,dataworld} systems to find candidate datasets or sellers. Data discovery search interfaces typically take natural language, datasets, or schemas as input. \sys uses existing data discovery systems~\cite{fernandez2018aurum} to identify vertical and horizontal augmentation candidates and is agnostic to the specific discovery mechanism.  

\stitle{Views} are commonly used for access control~\cite{Rizvi2004ExtendingQR}. They prevent access to the underlying data and only expose query results to the user. Since ML models can be expressed as aggregation queries, \sys uses ``ML Model'' views to protect seller datasets from release. 

Differentially private queries~\cite{Proserpio2014CalibratingDT} add noise to aggregated statistics to provide plausible deniability for individual input records. These techniques are compatible with \sys---noise can be added to the final model parameters, to the predictions, or potentially even by sellers before uploading their data to the service.  

\stitle{Factorized ML systems.} 
\revise{Many in-database ML works optimize model training over joins by implementing strategies such as introducing partitioning-preserving operations to optimize shuffle join~\cite{boehm2016systemml} or eliminating joins if their features are not used by the models~\cite{park2022end}. In contrast, factorized ML is an algorithmic optimization that translates ML into suitably designed semi-ring aggregations and pushes aggregations through joins and unions to achieve asymptotically lower time complexity, even for many-to-many joins. It supports many popular models (ridge regression~\cite{schleich2019layered}, SVM~\cite{khamis2020functional}, factorization machine~\cite{schleich2019layered}) and approximates others (k-means~\cite{curtin2020rk}, GLM~\cite{huggins2017pass}). However, previous factorized ML works optimize a single model training, while data search requires training numerous models. As shown in \Cref{exp:fac_learning}, \sys considerably accelerates factorized ML by preprocessing and sharing computations.}

%% file: sections/conclusions.tex
\section{Conclusions}
\label{sec:conclusions}

This paper presented \sys, a data-centric AutoML. 
Compared to existing AutoML,
\sys leverages the rich data corpus to augment training data with new features and examples.
To identify augmentation opportunities efficiently, \sys uses factorized learning with aggressive pre-computations, and a cost model for balancing the search cost and model training. Our experiments show that   \sys returns considerably more accurate models in orders of magnitude less time than SOTA open-source and commercial AutoML services.